\documentclass[11pt,a4paper]{article}

\usepackage[english]{babel}
\usepackage[utf8]{inputenc}

\usepackage{epsf}
\usepackage{cite}
\usepackage{graphicx}
\usepackage{subcaption} 

\usepackage{amsmath}
\usepackage{amssymb}
\usepackage{mathrsfs}
\usepackage[normalem]{ulem}
\usepackage{mathbbol}
\usepackage{multirow}
\usepackage{slashed}
\usepackage{amsfonts}
\usepackage{bbding}
\usepackage{array}
\usepackage[
colorlinks=true
,urlcolor=black
,anchorcolor=black
,citecolor=black
,filecolor=black
,linkcolor=black
,menucolor=black
,linktocpage=true
,pdfproducer=medialab
,pdfa=true
]{hyperref}

\usepackage{color} 
\usepackage[left=2cm,right=2cm,top=2cm,bottom=2cm]{geometry}
\usepackage{tikz}
\usetikzlibrary{shapes,arrows,positioning,automata,backgrounds,calc,er,patterns}
\usepackage[compat=1.1.0]{tikz-feynman}
\usepackage{contour}
\usepackage{placeins}
\allowdisplaybreaks
\begin{document}
\begin{center}
\vspace*{1mm}
\vspace{1.3cm}
{\Large\bf
Revisiting cLFV in ``T1-2-A'' scotogenic models: 
\vspace*{3mm}\\
asymmetries in three-body lepton decays}

\vspace*{1.2cm}
{\bf Adrian Darricau and  
Ana~M.~Teixeira }

\vspace*{.5cm}
Laboratoire de Physique de Clermont Auvergne (UMR 6533), CNRS/IN2P3,\\
Univ. Clermont Auvergne, 4 Av. Blaise Pascal, 63178 Aubi\`ere Cedex,
France
 \end{center}

\vspace*{5mm}
\begin{abstract}
\noindent
We consider a well-motivated class of scotogenic models (the ``T1-2-A'' variant), and carry out a comprehensive reassessment of its prospects regarding charged lepton flavour violating (cLFV) observables. 
Aiming only at explaining neutrino oscillation data and putting forward a viable dark matter candidate, a thorough exploration of the model's parameter space suggests that one can have sizeable rates for cLFV observables,
especially in rare muon transitions. 
We have further considered the role of parity and time-reversal asymmetries for cLFV 3-body decays,  $\ell_\alpha^+ \to \ell_\beta^+ \ell_\gamma^+ \ell_\delta^-$, which can be potentially studied in association with polarised muon and tau decays. The new set of observables offers further complementarity information on the scotogenic model under consideration, and possible means of testing it.

\end{abstract}

\section{Introduction}
Together with an explanation to the baryon asymmetry of the Universe (BAU), understanding the origin of neutrino mass generation and the observed dark matter (DM) relic density remain fundamental open problems in particle physics. Many well-motivated extensions of the Standard Model (SM) allow addressing the above mentioned observational issues, while rendering less severe certain theoretical caveats of the SM. 
Relying on a strong connection between viable dark matter candidates and a natural explanation to the smallness of the light neutrino masses, scotogenic models~\cite{Tao:1996vb,Ma:2006km,Fraser:2014yha} stand as a very interesting class of New Physics (NP) constructions. 

In recent years, numerous realisations have been investigated (see e.g.~\cite{Restrepo:2013aga}), among them the so-called ``T1-2-A'' scotogenic variants.
This class of NP models extends the SM via one doublet and a singlet scalar fields, with the fermion sector  comprising an additional Dirac doublet and two fermion singlets~\cite{Sarazin:2021nwo}. It offers extensive phenomenological advantages, as in addition to putting forward viable explanations to the three SM observational problems, it could accommodate the formerly existing tension in the anomalous magnetic moment of the muon~\cite{Aoyama:2020ynm}, and lead to extensive contributions to a variety of low-energy lepton observables, including charged lepton flavour violating (cLFV) transitions and decays~\cite{Alvarez:2023dzz}.

The intensive study of a given model's predictions regarding cLFV, in particular of peculiar features such as correlated behaviours between observables (as a consequence of the dominance of a given operator), has been explored in order to ascertain whether it can be at the origin of a cLFV observation in the future~\cite{Calibbi:2017uvl}. This has been the case of several scotogenic constructions (see, e.g.~\cite{Toma:2013zsa}), for which strong correlations between several cLFV observables offered means of falsifying specific realisations. 
However, and while certain of these ``cLFV-signatures''
do reflect an intrinsic property of the NP construction, others only hold under (simplifying) assumptions or then for specific regimes. 

Regarding the scotogenic ``T1-2-A'' realisation, 
previous cLFV studies focused on the implication of the dipole dominance regime which was favoured by an explanation of the former tension in the muon anomalous magnetic moment, $(g-2)_\mu$; this suggested that strong correlations occurred between radiative and dipole-mediated three-body decays (as well as neutrinoless muon-electron conversion in nuclei)~\cite{Alvarez:2023dzz, Darricau:2025vcs}. 
As the discrepancy between SM prediction and experimental observation for the muon anomalous magnetic moment ($\Delta a_\mu$)  became less significant, the correlation between cLFV observables became less pronounced, although still manifest~\cite{Darricau:2025vcs}. 

The new results regarding the SM prediction for $(g-2)_\mu$ have led to a striking change of paradigm: in view of the remarkable agreement between theory and experiment (under one part per million~\cite{Aliberti:2025beg, Muong-2:2025xyk}), the longstanding NP beacon now emerging as a SM electroweak (EW) precision observable. We have thus set to revisit the ``T1-2-A'' scotogenic model, and explore in a fully unbiased manner its potential regarding charged lepton flavour violation. In the present study not only we do relax requirements on accounting for the BAU via leptogenesis (and saturating the formerly existing tension in $(g-2)_\mu$), but we also rely on a new, more sophisticated scanning technique of the highly non-trivial parameter space. As we will subsequently discuss, the new survey will allow exploring new regions in the parameter space, which exhibit interesting features in what concerns cLFV observables. Significant contributions to leptonic processes can emerge in these new phenomenologically viable regimes (stemming from penguin and box diagrams), leading to cLFV rates within experimental sensitivity.  
More importantly, the departure from dipole dominance in cLFV three-body decays and $\mu-e$ conversion in nuclei does erode the pre-existing correlated behaviours between observables, which were powerful probes to falsify this class of models. 

Enlarging the set of cLFV observables that can be likely explored in dedicated experiments is thus paramount to probe (and falsify) NP scenarios as the present one. In recent years, asymmetries have been considered as a source of new observables, which are instrumental to disentangle the NP model as they potentially allow discriminating between the contributing operators. 
This has been the case of CP-violating asymmetries in cLFV $Z$ decays~\cite{Abada:2022asx} and more recently, of asymmetries in cLFV three-body decays~\cite{Goto:2010sn, Bolton:2022lrg,Redigolo:2024ztw, Darricau:2025rmu}. Given the highly polarised muon beams at PSI
(Mu3e experiment~\cite{Mu3e:2020gyw}), and potential strategies to obtain polarised tau-samples~\cite{Hagiwara:1989fn,Alemany:1991ki}, it might be possible to study the angular observables should a cLFV 3-body decay be discovered.
In what concerns the model under study, having new operators contributing to cLFV 3-body decays thus allows for potential interferences, implying that one can explore new observables, as $T$, $P$ and $P^\prime$ asymmetries.
While the new observables do not allow recovering the powerful probing power of cLFV correlation patterns, they nevertheless allow shedding new light on the ``T1-2-A'' realisation. 

The present manuscript is organised as follows: in Section~\ref{sec:model}, we begin by a short overview of the scotogenic model under consideration, including some highlights of a new exploration of its parameter space. Section~\ref{sec:clfv:amps:asym} is devoted to cLFV observables, including the presentation of the asymmetries in cLFV three-body decays. Our results are then presented in Section~\ref{sec:results}, and we summarise the most significant findings in the Outlook. Further details and complementary information can be found in the appendices

\section{The ``T1-2-A'' scotogenic model}\label{sec:model}

In this section we briefly describe the most important features of the model explored in~\cite{Alvarez:2023dzz}, and which has been recently revisited~\cite{Darricau:2025vcs}: we detail the new field content and interactions, outline the model's contributions to a variety of observables, and summarise the main results of our survey of the parameter space.

\subsection{Brief description of the model}
The so-called ``T1-2-A'' variant consists in the addition of several new fields to the SM content, in particular
an SU(2)$_L$ scalar doublet and a real scalar singlet (respectively, $\eta$ and $S$), two Majorana fermion singlets ($F_{1,2}$) and finally vector-like Dirac fermions, which are doublets under SU(2)$_L$, $\Psi_{1,2}$. The associated SU(2)$_L \times$U(1)$_Y$ charges are given in Table~\ref{table:NPcontent}. All NP fields are odd under the discrete $Z_2$ symmetry, enforced to ensure the stability of the lightest neutral state of the new spectrum which is then a potential DM candidate; conversely all the SM fields are even under the latter symmetry (further notice that only $Z_2$-even fields can carry lepton number).
\begin{table}[h!]
\centering\renewcommand{\arraystretch}{1.4} 
\begin{tabular}{c|cc|cccc}
\hline
\hline
Field & $\eta$ & $S$ & $F_{1}$ & $F_{2}$ & $\Psi_{1}$ & $\Psi_{2}$\\ 
\hline
SU$(2)_L$ & $\mathbf{2}$ & $\mathbf{1}$ & $\mathbf{1}$ & $\mathbf{1}$ & $\mathbf{2}$ & $\mathbf{2}$ \\
U$(1)_Y$ & $1$ & $0$ & $0$ & $0$ & $-1$ & $1$  \\
\hline
\hline
\end{tabular}
\caption{Additional field content of the ``T1-2-A'' scotogenic model variant (cf.~\cite{Alvarez:2023dzz}). All the new fields are odd under the $Z_2$ symmetry.} 
\label{table:NPcontent}
\renewcommand{\arraystretch}{1} 
\end{table}

In view of the new fields and associated interactions, 
the SM Lagrangian is extended as
\begin{align}\label{eq:lagrangian:fermion}
\mathcal{L}_\text{fermion}\, &= i\, (\overline{\psi_i} \,\gamma^\mu\, D_\mu\, \psi_i + \overline{F_i} \,\gamma^\mu \,D_\mu \,F_i) \nonumber \\ 
&- M_\psi \,\bar{\psi_1} \,\tilde{\psi_2} - \frac{1}{2} \,{M_F}_{ii} \,\overline{F_i^c}\, F_i + y_{1 i}^*\, \overline{F_i} \,\Phi^\dagger \,\tilde{\psi_1} + y_{2 i}^* \,\overline{F_i} \,\Phi \,\psi_2^c \nonumber\\
&- g_\psi^\alpha \,\bar{\tilde{\psi_2}} \,L^\alpha_L S - g_{F_i}^\alpha \,\overline{\widetilde{L_L^\alpha}}\, \eta F_i - g_R^\alpha \overline{e_R^\alpha} \,\eta^\dagger \,\Psi_1 + \text{H.c.}\, .
\end{align}
In the above, we have denoted the fermion doublets as $\Psi_1 = (\Psi_1^0, \Psi_1^-)^T$ and $\Psi_2 = ({\Psi_2^-}^c, -{\Psi_2^0}^c)^T$; regarding the SM fields, $L$ and $e^c$ correspond to the left- and right-handed lepton multiplets, with $\tilde H = i \sigma_2 H^*$ (similarly for $\eta$) and $\tilde{\psi} = i \sigma_2 \psi^c$. Notice that $g_\Psi$, $g_F$ and $y_i$ violate lepton number conservation, and are thus expected to enter directly in the mechanism of neutrino mass generation (as subsequently discussed).
After electroweak symmetry breaking (EWSB), the fermion spectrum further comprises\footnote{As usually done, in the present work, greek indices correspond to lepton flavours ($\alpha, \beta=e, \mu, \tau$), while in Eq.~(\ref{eq:lagrangian:fermion}) $i,j=1,2$ denote generations of fields; after EWSB latin indices generically denote mass eigenstates.} charged heavy Dirac fermions, $\Psi^\pm$, whose masses are trivially given 
by $M_\Psi$, and four Majorana neutral fermions, $\chi^0_i$, resulting from the mixings of 
$F_1$, $F_2$, $\Psi^0_1$, and ($\Psi^0_2)^c$. The full mass matrices and the associated diagonalisation procedure can be found in Appendix~\ref{app:model:description}. 

Concerning the scalar sector, new terms are present:
\begin{align}\label{eq:Vscalar}
\mathcal{V}_\text{scalar}\, &= \frac{1}{2} M_S^2\, S^2 + \frac{1}{2} \lambda_{4 S} \,S^4 + M_\eta^2 \,|\eta|^2 + \lambda_{4 \eta} \,|\eta|^4 + \frac{1}{2} \lambda_S \,S^2 |H|^2 + \frac{1}{2} \lambda_{S \eta} \,S^2 |\eta|^2 \nonumber \\ &
+ \lambda_{\eta} \,|\eta|^2\,|H|^2 + \lambda_{\eta}^\prime \,|\eta H^\dagger|^2 + \frac{1}{2} \lambda_{\eta}^{\prime \prime} \left[ \left(H \eta^\dagger \right)^2 +\text{H.c.} \right] + \alpha \,S \left[ H \eta^\dagger + \text{H.c.} \right].
\end{align}
The physical scalar spectrum is thus composed of the SM Higgs, additional neutral states (two scalars and one pseudoscalar - $\phi_1$, $\phi_2$ and $A_0$) and a charged scalar ($\eta^\pm$); the mass matrices and relations between interaction and mass bases are also summarised in Appendix~\ref{app:model:description}. 

\subsection{Phenomenological impact}\label{sec:pheno_impact}
By construction, this class of SM extensions aims at simultaneously accommodating neutrino oscillation phenomena and complying with the currently observed relic dark matter abundance (and associated detection constraints). Moreover, the new states and their interactions with the SM fields will be at the origin of extensive contributions to a wide variety of observables: transitions and decays already present in the SM (as is the case of EW precision observables, or observables sensitive to deviations from lepton flavour universality, as discussed in~\cite{Darricau:2025vcs}); processes entirely forbidden in the SM, such as cLFV rare transitions and decays, which are the focus of the present study.

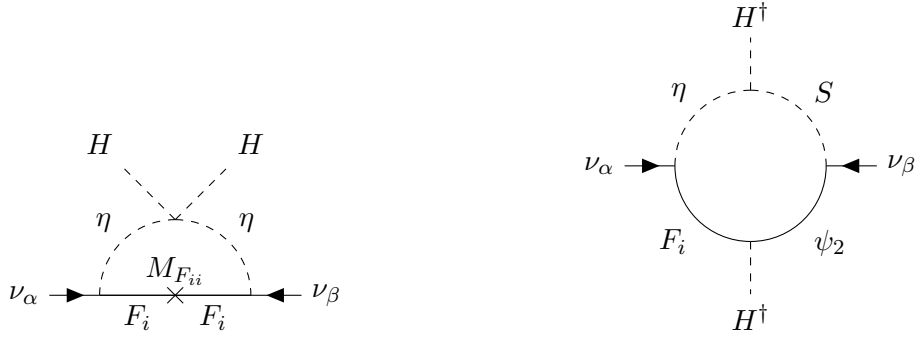
\begin{figure}[h!]
    \centering
    \begin{subfigure}[b]{0.48\textwidth}
        \centering
        \begin{tikzpicture}
        \begin{feynman}
        \vertex (a) at (0,0) {\(\nu_\alpha\)};
        \vertex (b) at (1,0);
        \vertex (c) at (2,0);
        \vertex (d) at (2,1);
        \vertex (e) at (3,0);
        \vertex (f) at (4,0) {\(\nu_\beta\)};
        \vertex (g) at (3,2) {\( H\)};
        \vertex (h) at (1,2) {\( H\)};
        \diagram* {
        (a) -- [fermion] (b),
        (b) -- [plain, edge label'=\( F_i\)] (c),
        (c) -- [plain, edge label'=\( F_i\)] (e),
        (b) -- [scalar, quarter left, edge label=\( \eta\)] (d),
        (d) -- [scalar, quarter left, edge label=\( \eta\)] (e),
        (b) -- [insertion = 0.5, edge label=\( M_{F_{ii}}\)] (e),
        (f) -- [fermion] (e),
        (d) -- [scalar] (g),
        (d) -- [scalar] (h)
        };
        \end{feynman}
        \end{tikzpicture}
    \end{subfigure}
    \begin{subfigure}[b]{0.4\textwidth}
        \centering
        \begin{tikzpicture}
        \begin{feynman}
        \vertex (a) at (0,0) {\(\nu_\alpha\)};
        \vertex (b) at (1,0);
        \vertex (c) at (2,-1);
        \vertex (d) at (2,1);
        \vertex (e) at (3,0);
        \vertex (f) at (4,0) {\(\nu_\beta\)};
        \vertex (g) at (2,2) {\( H^\dagger\)};
        \vertex (h) at (2,-2) {\( H^\dagger\)};
        \diagram* {
        (a) -- [fermion] (b),
        (b) -- [plain, quarter right, edge label'=\( F_i\)] (c),
        (c) -- [plain, quarter right, edge label'=\( \psi_2\)] (e),
        (b) -- [scalar, quarter left, edge label=\( \eta\)] (d),
        (d) -- [scalar, quarter left, edge label=\( S\)] (e),
        (f) -- [fermion] (e),
        (d) -- [scalar] (g),
        (c) -- [scalar] (h)
        };
        \end{feynman}
        \end{tikzpicture}
    \end{subfigure}
    \caption{One-loop diagrams contributing to neutrino masses (in the interaction basis).}
    \label{fig:NMassInt}
\end{figure}

\paragraph{Neutrino mass generation}
As is the case in scotogenic realisations, loop-level neutrino mass generation\footnote{Recall that the discrete symmetry ensuring the stability of the potential DM candidate precludes tree-level contributions to neutrino masses.}, together with the smallness of the new couplings allows for a more natural explanation of neutrino masses and oscillation data. In the present ``T1-2-A'' scotogenic variant, the one-loop contributions to Majorana neutrino masses are schematically depicted in Fig.~\ref{fig:NMassInt} (in the interaction basis).
These diagrams are the source of non-vanishing 
contributions to the neutrino masses of the form  
$\overline{\nu_{\beta}^{c}} \left( \mathcal{M}_{\nu} \right)_{\beta \alpha} \nu_{\alpha}$.  After electroweak symmetry breaking, the neutrino mass matrix ($\mathcal{M}_{\nu}$) can be written in terms of a generalised matrix of ``couplings''
$\mathcal{G}$, and the associated loop-contributions, $\mathcal{M}_{L}$~\cite{Alvarez:2023dzz}:
\begin{equation}\label{eq:Gmatrix}
    \mathcal{M}_{\nu} = \mathcal{G}^T \,\mathcal{M}_{L} \,\mathcal{G}, \quad \text{with} \quad \mathcal{G} = \begin{pmatrix}
        g_{\psi}^{e} & g_{\psi}^{\mu} & g_{\psi}^{\tau} \\[5pt]
        g_{F_{1}}^{e} & g_{F_{1}}^{\mu} & g_{F_{1}}^{\tau} \\[5pt]
        g_{F_{2}}^{e} & g_{F_{2}}^{\mu} & g_{F_{2}}^{\tau}
    \end{pmatrix}\,,
\end{equation}
in which $\mathcal{M}_{L}$ is given in Appendix~\ref{app:model:description}. Compliance with neutrino oscillation data can be 
achieved by means of a modified Casas-Ibarra parametrisation of 
$\mathcal{G}$ (via $g_\psi$ and $g_F$)~\cite{Casas:2001sr,Basso:2012voo} 
\begin{equation}\label{eq:CI-parametrisation}
    \mathcal{G} \,= U_{L} \,D_{L}^{-1/2}\, R\, D_{\nu}^{1/2} \,U_{\text{PMNS}}^{*}\,,
\quad \text{with} \quad 
    D_{L} \,= \,U_{L}^{T} \,\mathcal{M}_{L} \,U_{L}\,,
\end{equation}
in which  $D_{\nu}$ is the diagonal matrix of the physical (light) neutrino masses, and with the unitary $3 \times 3$ $U_{\text{PMNS}}$ matrix encoding leptonic mixings; the $R$ mixing matrix contains the remaining degrees of freedom.
Regarding neutrino oscillation data, we take the most recent results from the NuFit collaboration~\cite{Esteban:2024eli}, and consider a normal ordering of the light neutrino spectrum.\footnote{The inverse ordering leads to a similar phenomenology; for simplicity, we only consider here a normal ordering of the light neutrino spectrum.}
Further details are provided in Appendix~\ref{app:model:description}.

\paragraph{Viable dark matter candidate}\label{par:DM}
In the present scotogenic realisation, viable DM candidates emerge in the form of 
the ($Z_{2}$-odd) lightest state of the extended neutral fermion and scalar sectors. The DM candidate can be the lightest CP-even scalar $\phi_{1}$, the CP-odd scalar $A^{0}$ or the lightest fermion $\chi_{1}^{0}$. 
All the aspects of DM phenomenology (computation of the relic density and compliance with direct and indirect detection bounds) have been evaluated via 
{micrOMEGAs}~\cite{Alguero:2023zol}. In particular, we require the relic density to lie within the $3 \sigma$ bound from theoretical uncertainties,
\begin{equation}\label{eq:omega:planck}
    \Omega_{\operatorname{CDM}} \,h^2 \,= \,0.120 \pm 0.012\,,
\end{equation}
further imposing that the spin-independent direct detection cross-section be below the LUX-ZEPLIN~\cite{LZ:2022lsv} experimental limit (at 95\% C.L.).

\paragraph{Electroweak precision observables}
In our study we take into account several of EW precision observables. These include flavour conserving leptonic $Z$ and $H$ decays (and ratios of individual rates as probes of lepton flavour universality violation - LFUV), invisible $Z$ and $H$ decays, as well as oblique parameters.  For a full description of the computation of these observables, including details on the renormalisation of the interaction vertices, and discussion of their impact in constraining the model's parameter space, we refer to~\cite{Darricau:2025vcs}, in which a thorough study was first carried out. In Table~\ref{tab:obs:EW}, we summarise the SM predictions and the current experimental status for a subset of these observables.
\renewcommand{\arraystretch}{1.3}
\begin{table}[h!]
    \centering
    \hspace*{-2mm}{\small\begin{tabular}{|c|c|c|}
    \hline
    Observable & Exp. measurement & SM prediction  \\
    \hline
    $\Gamma(Z\to e^+e^-)$ & $83.91\pm0.12\:\mathrm{MeV}$ (LEP~\cite{ALEPH:2005ab}) & $83.965\pm0.016\:\mathrm{MeV}$~\cite{Freitas:2014hra}\\
    $\Gamma(Z\to \mu^+\mu^-)$ & $83.99\pm0.18\:\mathrm{MeV}$ (LEP~\cite{ALEPH:2005ab}) & $83.965\pm0.016\:\mathrm{MeV}$~\cite{Freitas:2014hra}\\
    $\Gamma(Z\to \tau^+\tau^-)$ & $84.08\pm0.22\:\mathrm{MeV}$ (LEP~\cite{ALEPH:2005ab}) & $83.775\pm0.016\:\mathrm{MeV}$~\cite{Freitas:2014hra}\\
    \hline
    $\Gamma(Z\to\mathrm{inv.})$ & $499.0 \pm 1.5\:\mathrm{MeV}$ (PDG~\cite{ParticleDataGroup:2024cfk})& $501.45\pm 0.05\:\mathrm{MeV}$~\cite{Freitas:2014hra}\\
    \hline
    $R_{\mu e}(Z\to\ell\ell)$ & $1.0001\pm 0.0024$ (PDG~\cite{ParticleDataGroup:2024cfk}) & $1.0$~\cite{Freitas:2014hra}\\
    $R_{\tau e}(Z\to\ell\ell)$ & $1.0020\pm0.0032$ (PDG~\cite{ParticleDataGroup:2024cfk}) & $0.9977$~\cite{Freitas:2014hra}\\
    $R_{\tau \mu}(Z\to\ell\ell)$ & $1.0010\pm 0.0026$ (PDG~\cite{ParticleDataGroup:2024cfk}) & $0.9977$~\cite{Freitas:2014hra}\\
    \hline
    $R_{\tau \mu}(H\to\ell\ell)$ & $230\pm 146$ (PDG~\cite{ParticleDataGroup:2024cfk}) & $288$~\cite{LHCHiggsCrossSectionWorkingGroup:2016ypw}\\
    \hline
    $\mathrm{BR}(H\to\tau^+\tau^-)$ & $0.06_{-0.007}^{+0.008}$ (PDG~\cite{ParticleDataGroup:2024cfk}) & $0.0624\pm0.0035$~\cite{Denner:2011mq}\\
    $\mathrm{BR}(H\to\mu^+\mu^-)$ & $(2.6 \pm 1.3)\times 10^{-4}$ (PDG~\cite{ParticleDataGroup:2024cfk}) & $(2.17 \pm0.13)\times 10^{-4}$~\cite{Denner:2011mq}\\
    \hline
    \end{tabular}}
    \caption{Experimental  values and SM predictions  for several LFUV and EW observables. All uncertainties are given at 68\% C.L. (the parametric uncertainties are negligible for the SM predictions of the universality ratios).}
    \label{tab:obs:EW}
\end{table}
\renewcommand{\arraystretch}{1.}

\medskip
Let us also mention that in our analysis - and in view of the fact that there are several new sources of CP violation in the present model - we systematically ensured compatibility with the constraint emerging from the experimental bound on the electron electric dipole moment~\cite{ParticleDataGroup:2024cfk},
\begin{equation}\label{eq:eEDM}
       \vert d_e \vert < 4.1 \times 10^{-30}\, e.\text{cm}\,.
\end{equation}

\paragraph{The role of $\pmb{(g-2)_\mu}$}
In the past decades, emphasis was put on a NP model's capacity to saturate the discrepancy between the SM prediction and observation regarding $(g-2)_\mu$, a tension that neared the $5\sigma$ level. 
However, the most recent combined theory predictions~\cite{Aliberti:2025beg} reveal an excellent agreement with the Fermilab Muon $g-2$ collaboration world average~\cite{Muong-2:2025xyk}. The discrepancy between theory and experiment for the anomalous magnetic moment of the muon, $\Delta a_\mu$, now stands at 
\begin{equation}\label{eq:ammmu:delta}
    \Delta a_\mu \,= \,(3.9 \pm 6.4) \times 10^{-10}\,,
\end{equation}
rendering it a powerful test of the SM.

\subsection{Exploring the scotogenic parameter space}

In what follows we briefly overview how the parameter space of the model was explored, building upon
the approach of recent studies~\cite{deSouza:2025uxb,Darricau:2025vcs,Alvarez:2023dzz}. 

In view of the formerly existing tension concerning the anomalous magnetic moment of the muon, previous studies of the ``T1-2-A'' scotogenic realisation put a strong emphasis on saturating the  deviation between the SM prediction and observation. In order to do so while avoiding excessive  contributions to cLFV observables, a
special hierarchy in the upper row of the $\mathcal{G}$ matrix was required. 
In the present study, and instead of indirectly having certain flavour observables as inputs of a global scan, all observables (flavoured rates and ratios, electric and magnetic moments) are now strictly outputs. Only neutrino oscillation data is treated as an input via the Casas-Ibarra parametrisation.
Building up from the input parameters, which are summarised in Table~\ref{tab:Input_parameters} (and closely following the recent results of~\cite{deSouza:2025uxb}), we implemented a  differential evolution Markov-Chain Monte Carlo (DE-MCMC) algorithm.
The $\chi^2$ distribution was constructed aiming at following the  experimental constraints previously 
discussed in Section~\ref{sec:pheno_impact}. Additional weights were assigned to points leading to $\ell \to 3 \ell^{\prime}$ decay rates within future experimental  sensitivities. Further bonus weights were allocated to 
points in parameter space which satisfied all experimental constraints.

\begin{table}[h!]
\centering\renewcommand{\arraystretch}{1.8} 
\begin{tabular}{cc|cc}
\hline
\hline
\textbf{Parameter} & \textbf{Range} & \textbf{Parameter} & \textbf{Range} \\ 
\hline
$\alpha$ & $\pm\left[ 10^{-2}, 10^{4} \right]$ GeV & $M_{S}^{2}, M_{\eta}^{2}$ & $\left[ 5 \times 10^{5}, 5 \times 10^{6} \right]$ $\text{GeV}^2$ \\
$\lambda_{4S}, \lambda_{4\eta}, \lambda_{S \eta}, \lambda_S$ & $\pm \left[ 10^{-10}, 1 \right]$ & $M_{1}, M_{2}$ & $\left[ 100, 20000 \right]$ GeV \\ 
$\lambda_{\eta}, \lambda_{\eta^{\prime}}, \lambda_{\eta^{\prime\prime}}$ & $\pm\left[ 10^{-10}, 1 \right]$ & $M_{\psi}$ & $\left[ 700, 2000 \right]$ GeV \\
$m_{\nu_{1}}$ & $\left[ 10^{-19}, 10^{-10} \right]$ GeV& $y_{11, 12, 21, 22}$ & $\pm\left[ 10^{-10}, 1 \right]$ \\
$\vert \theta_R^i \vert$ & $\left[ 10^{-8}, 10^{3} \right]$ & $\vert g_R^\alpha \vert$ & $\left[ 10^{-10}, \sqrt{4 \pi}\right]$ \\
$\arg \theta_R^i$ & $[ 0, 2 \pi [$ & $\arg g_R^\alpha$ & $[ 0, 2 \pi [$
\\
\hline
\hline
\end{tabular}
\caption{Input parameter ranges used in the DE-MCMC scan.} 
\label{tab:Input_parameters}
\renewcommand{\arraystretch}{1} 
\end{table}

\begin{figure}[h!]
    \centering
    \includegraphics[width=0.45 \textwidth] {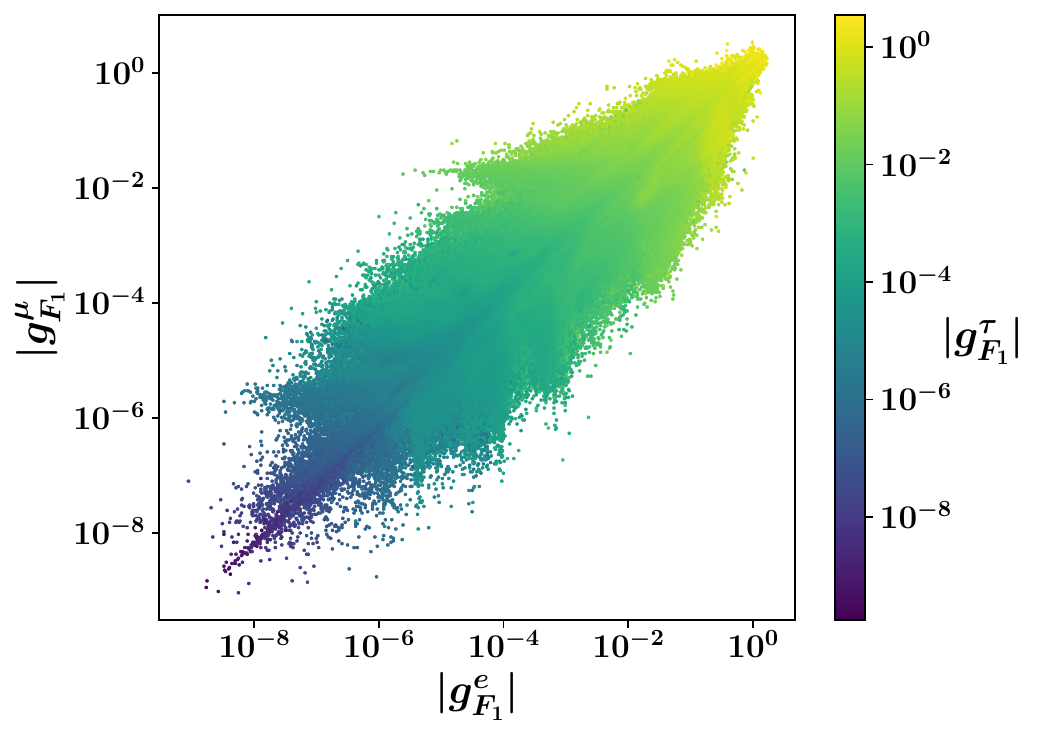}
    \hspace*{10mm}
    \includegraphics[width=0.45 \textwidth] {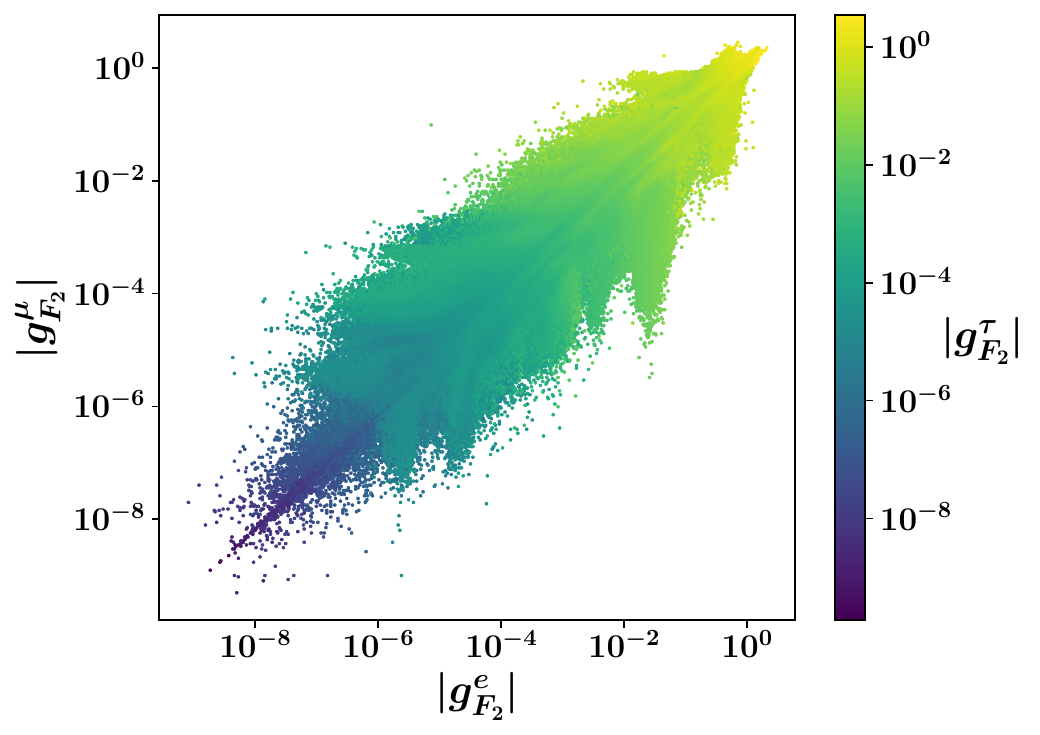}
    \\
    \includegraphics[width=0.45 \textwidth]{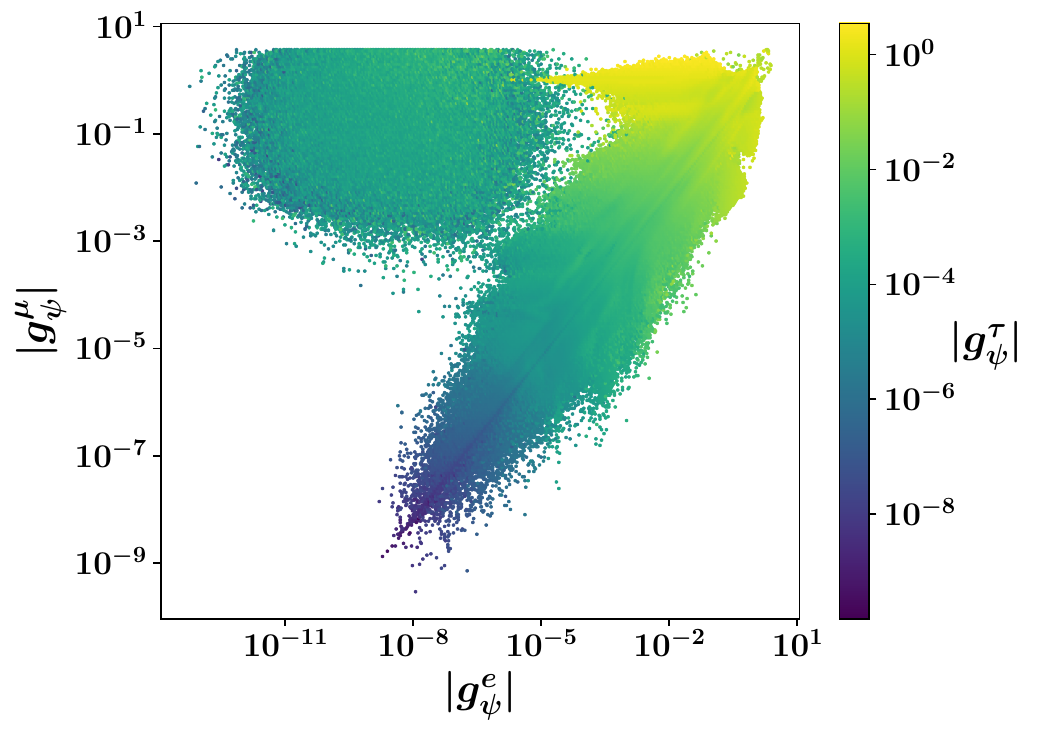}
    \hspace*{10mm}
    \includegraphics[width=0.45 \textwidth]{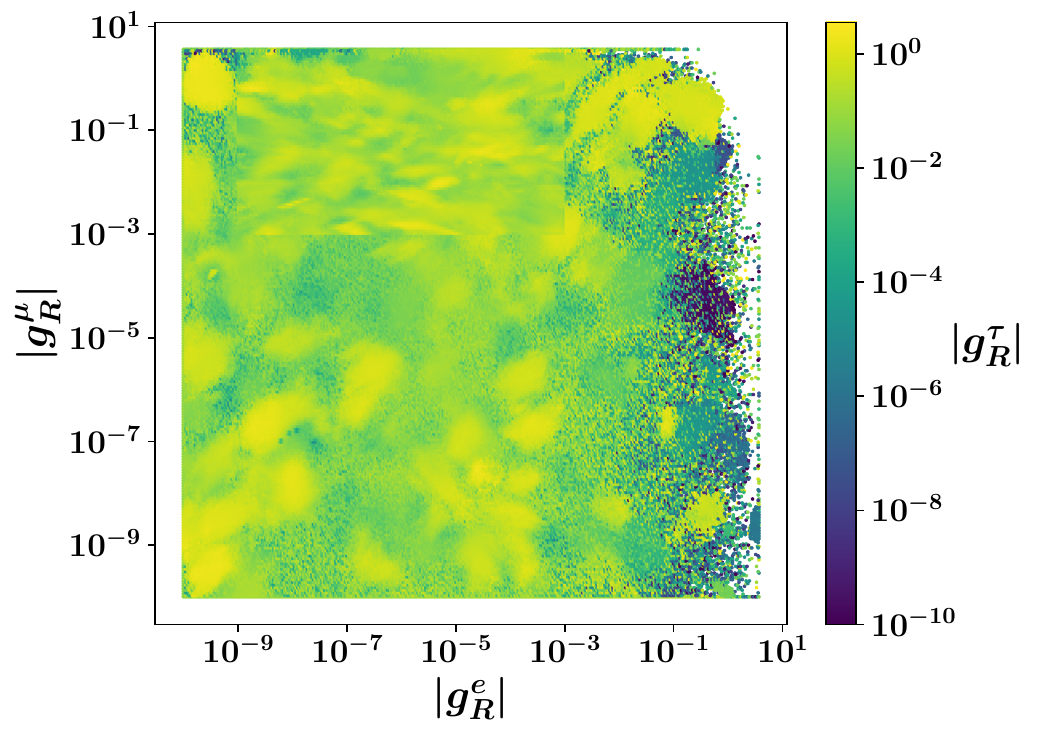}
    \caption{Absolute values of the couplings $g_{F_1}^\alpha$ (upper left), $g_{F_2}^\alpha$ (upper right), $g_{\psi}^\alpha$ (lower left), $g_{R}^\alpha$ (lower right) for $\alpha=e,\mu$, as obtained from the dedicated DE-MCMC survey of the parameter space. In all plots the colour palette is associated with the variation of the third generation coupling ($\alpha=\tau$).
    }\label{fig:Gmatrix}
\end{figure}

For completeness, we have carried out a comparative study of the (new) allowed parameter space for the ``T1-2-A'' scotogenic realisation. A summary is presented in Fig.~\ref{fig:Gmatrix}, in which we display $g_\psi^\alpha$ and $g_{F_i}^\alpha$ (which enter the definition of the generalised matrix of couplings, $\mathcal{G}$) and $g_{R}^\alpha$.

While the distributions for the $g_{F_i}^\alpha$ couplings are in good agreement with previous findings (cf.~\cite{deSouza:2025uxb,Darricau:2025vcs,Alvarez:2023dzz}), new regimes clearly emerge for $g_{\psi}^\alpha$ and $g_{R}^\alpha$ (lower panels of Fig.~\ref{fig:Gmatrix}). Regarding $g_{\psi}^\alpha$, one recovers the formerly observed region, corresponding to sizeable $g_{\psi}^\mu$ in association with small $g_{\psi}^{e,\tau}$; the latter 
stems from having the $\theta_R$ angles in the Casas-Ibarra parametrisation (see Eq.~(\ref{eqn:casas_ibarra})) highly correlated and fine-tuned. 
Interestingly, a new regime is also present, displaying a similar correlation pattern to those encountered for $g_{F_i}^\alpha$: such a regime corresponds to a more ``natural'' $\mathcal{G}$ matrix in association with uncorrelated $\theta_R$ angles. 
For the $g_{R}^\alpha$ couplings, and in contrast to previous findings, the plane is now densely filled , again a consequence of the less constrained $\theta_R$.

Throughout the phenomenologically allowed regimes (including reproducing the correct dark matter relic density), we notice that the spectrum of the NP states lies in the following intervals\footnote{Although a full study of LHC phenomenology clearly lies beyond the scope of this work, we have verified that our viable regimes are compatible with the bounds inferred in~\cite{deSouza:2025uxb}. In particular, our spectrum never exhibits $M_\psi \lesssim 900$~GeV in association with a considerable mass difference between the charged fermionic state and the DM singlet fermionic state ($M_\psi - M_{\chi_{_1}} \gtrsim 250$~GeV), which would be likely excluded by supersymmetry searches at the LHC (charginos and neutralinos), through $\psi^+ \to \chi_1^0 W^+$~\cite{ATLAS:2024qmx}. }
\begin{align}
 &   410~\text{GeV} \leq M_{\phi_1} \leq 2240~\text{GeV}\,, \quad 
    720~\text{GeV} \leq M_{\phi_2} \leq 2700~\text{GeV}\,, \nonumber \\
  &  690~\text{GeV} \leq M_{A^0} \leq 2250~\text{GeV}\,, \quad 
    700~\text{GeV} \leq M_{\eta^\pm} \leq 2240~\text{GeV}\,, \nonumber \\
 &    360~\text{GeV} \leq M_{\chi_{_1}} \leq 2000~\text{GeV}\,, \quad 
530~\text{GeV} \leq M_{\chi_{_2}} \leq 2000~\text{GeV}\,, \quad 
700~\text{GeV} \leq M_{\chi_{_{3,4}}} \leq 20~\text{TeV}\,.
\end{align}

On a secondary note, we have found that all three types of 
lightest $Z_2$-odd particles (fermion, scalar and pseudoscalar) can be viable DM candidates throughout the otherwise phenomenologically viable parameter space. Albeit not statistically significant, we do indeed identify regimes in which one can have a viable pseudoscalar DM candidate, absent from a previous study~\cite{Alvarez:2023dzz}.

As one can expect, the above mentioned differences in the allowed regimes - a consequence of the improved scanning procedure - will have a non-negligible impact for the prospects regarding cLFV observables, which we proceed to discuss.

\section{cLFV processes and asymmetries in 3-body decays}\label{sec:clfv:amps:asym}

Numerous extensions of the SM - in particular those incorporating a mechanism for neutrino mass generation - lead to non-negligible contributions to cLFV processes. 
Enlarging the set of cLFV observables is paramount to better disentangle the NP at the source of such transitions, especially in the presence of (new) sources of leptonic CP violation, as is the case of the present ``T1-2-A'' scotogenic realisation. 
In what follows, we first discuss contributions to cLFV muon and tau decays, and then present the different asymmetries that can be considered.

\subsection{Revisiting cLFV leptonic transitions in the ``T1-2-A'' scotogenic variant}\label{sec:clfv:revisit}

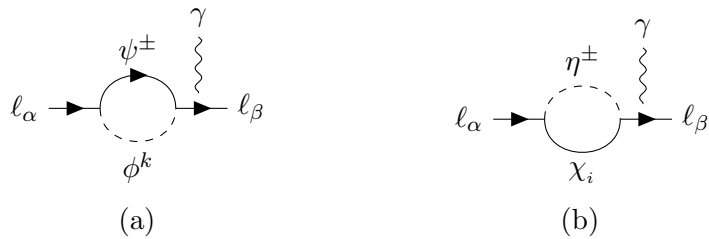
\begin{figure}[h!]
    \centering
        \begin{tikzpicture}
        \begin{feynman}
        \vertex (a) at (0,0) {\(\ell_\alpha\)};
        \vertex (b) at (1,0);
        \vertex (c) at (2,0);
        \vertex (d) at (3,0) {\(\ell_\beta\)};
        \vertex (e) at (2.3,0.2);
        \vertex (f) at (2.3,1.2) {\(\gamma\)};
        \diagram* {
        (a) -- [fermion] (b),
        (b) -- [fermion, half left, edge label=\( \psi^\pm\)] (c),
        (b) -- [scalar, half right, edge label'=\( \phi^k\)] (c),
        (c) -- [fermion] (d),
        (e) -- [boson] (f)
        };
        \end{feynman}
        \end{tikzpicture}
   \hspace*{20mm}
        \begin{tikzpicture}
        \begin{feynman}
        \vertex (a) at (0,0) {\(\ell_\alpha\)};
        \vertex (b) at (1,0);
        \vertex (c) at (2,0);
        \vertex (d) at (3,0) {\(\ell_\beta\)};
        \vertex (e) at (2.3,0.2);
        \vertex (f) at (2.3,1.2) {\(\gamma\)};
        \diagram* {
        (a) -- [fermion] (b),
        (b) -- [plain, half right, edge label'=\( \chi_{_i}\)] (c),
        (b) -- [scalar, half left, edge label=\( \eta^\pm\)] (c),
        (c) -- [fermion] (d),
        (e) -- [boson] (f)
        };
        \end{feynman}
        \end{tikzpicture}
        \mbox{\hspace*{31mm}
        (a)
        \hspace*{51 mm}
    (b)
    \hspace*{30mm}
        }
    \caption{Lepton-flavour violating dipole interactions for $\alpha \neq \beta$. For cLFV 3-body decays, the photon line is attached to a pair of opposite charge same flavour charged leptons ($\ell_\gamma^\pm$).}
    \label{fig:scoto:radiative:AMM:cLFV}
\end{figure}

\begin{figure}[h!]
    \centering
        \centering
        \begin{tikzpicture}
        \begin{feynman}
        \vertex (a) at (0,0) {\(Z\)};
        \vertex (b) at (1,0);
        \vertex (c) at (2,-1);
        \vertex (d) at (2,1);
        \vertex (e) at (3,-1) {\(\ell_\beta\)};
        \vertex (f) at (3,1) {\(\ell_\alpha\)};
        \diagram* {
        (a) -- [boson] (b),
        (b) -- [plain, edge label'=\( \chi_{_i}\)] (c),
        (b) -- [plain, edge label=\( \chi_{_j}\)] (d),
        (c) -- [scalar, edge label'=\( \eta^\pm\)] (d),
        (d) -- [fermion] (f),
        (e) -- [fermion] (c)
        };
        \end{feynman}
        \end{tikzpicture}
    \hfill
        \centering
        \begin{tikzpicture}
        \begin{feynman}
        \vertex (a) at (0,0) {\(Z\)};
        \vertex (b) at (1,0);
        \vertex (c) at (2,-1);
        \vertex (d) at (2,1);
        \vertex (e) at (3,-1) {\(\ell_\beta\)};
        \vertex (f) at (3,1) {\(\ell_\alpha\)};
        \diagram* {
        (a) -- [boson] (b),
        (b) -- [scalar, edge label'=\( \phi_{_k}\)] (c),
        (b) -- [scalar, edge label=\( \phi_l\)] (d),
        (c) -- [fermion, edge label'=\( \psi^\pm\)] (d),
        (d) -- [fermion] (f),
        (e) -- [fermion] (c)
        };
        \end{feynman}
        \end{tikzpicture}
    \hfill
        \centering
        \begin{tikzpicture}
        \begin{feynman}
        \vertex (a) at (0,0) {\(Z\)};
        \vertex (b) at (1,0);
        \vertex (c) at (2,-1);
        \vertex (d) at (2,1);
        \vertex (e) at (3,-1) {\(\ell_\beta\)};
        \vertex (f) at (3,1) {\(\ell_\alpha\)};
        \diagram* {
        (a) -- [boson] (b),
        (c) -- [fermion, edge label=\( \psi^+\)] (b),
        (b) -- [fermion, edge label=\( \psi^-\)] (d),
        (c) -- [scalar, edge label'=\( \phi_{_k}\)] (d),
        (d) -- [fermion] (f),
        (e) -- [fermion] (c)
        };
        \end{feynman}
        \end{tikzpicture}
    \hfill
        \centering
        \begin{tikzpicture}
        \begin{feynman}
        \vertex (a) at (0,0) {\(Z\)};
        \vertex (b) at (1,0);
        \vertex (c) at (2,-1);
        \vertex (d) at (2,1);
        \vertex (e) at (3,-1) {\(\ell_\beta\)};
        \vertex (f) at (3,1) {\(\ell_\alpha\)};
        \diagram* {
        (a) -- [boson] (b),
        (c) -- [scalar, edge label=\( \eta^+\)] (b),
        (b) -- [scalar, edge label=\( \eta^-\)] (d),
        (c) -- [plain, edge label'=\( \chi_{_i}\)] (d),
        (d) -- [fermion] (f),
        (e) -- [fermion] (c)
        };
        \end{feynman}
        \end{tikzpicture}
\mbox{  \hspace*{19mm}
            (a) \hspace*{38mm}
            (b) \hspace*{36mm}
            (c) \hspace*{36mm}
            (d) \hspace*{25mm}
    }
    \\
        \centering
        \begin{tikzpicture}
        \begin{feynman}
        \vertex (a) at (0,0) {\(Z\)};
        \vertex (b) at (1,0);
        \vertex (c) at (3,1) {\(\ell_\alpha\)};
        \vertex (d) at (1.5,-0.3);
        \vertex (e) at (2.2,-0.6);
        \vertex (f) at (3,-1) {\(\ell_\beta\)};
        \diagram* {
        (a) -- [boson] (b),
        (b) -- [fermion] (c),
        (d) -- [plain, half right, edge label'=\( \chi_{_i}\)] (e),
        (d) -- [scalar, half left, edge label=\( \eta^\pm\)] (e),
        (d) -- [fermion] (b),
        (f) -- [fermion] (e)
        };
        \end{feynman}
        \end{tikzpicture}
        \hspace*{7mm}
        \begin{tikzpicture}
        \begin{feynman}
        \vertex (a) at (0,0) {\(Z\)};
        \vertex (b) at (1,0);
        \vertex (c) at (3,1) {\(\ell_\alpha\)};
        \vertex (d) at (1.5,-0.3);
        \vertex (e) at (2.2,-0.6);
        \vertex (f) at (3,-1) {\(\ell_\beta\)};
        \diagram* {
        (a) -- [boson] (b),
        (b) -- [fermion] (c),
        (e) -- [fermion, half left, edge label=\( \psi^\pm\)] (d),
        (e) -- [scalar, half right, edge label'=\( \phi_{_k}\)] (d),
        (d) -- [fermion] (b),
        (f) -- [fermion] (e)
        };
        \end{feynman}
        \end{tikzpicture}\\
    \mbox{  \hspace*{12mm}
            (e) \hspace*{36mm}
            (f)
    }
    \caption{Lepton-flavour violating $Z$ interactions for $\alpha \neq \beta$. For cLFV 3-body decays, the vector line is attached to a pair of opposite charge same flavour charged leptons ($\ell_\gamma^\pm$). 
    }
    \label{fig:scoto:Zlldiag}
\end{figure}
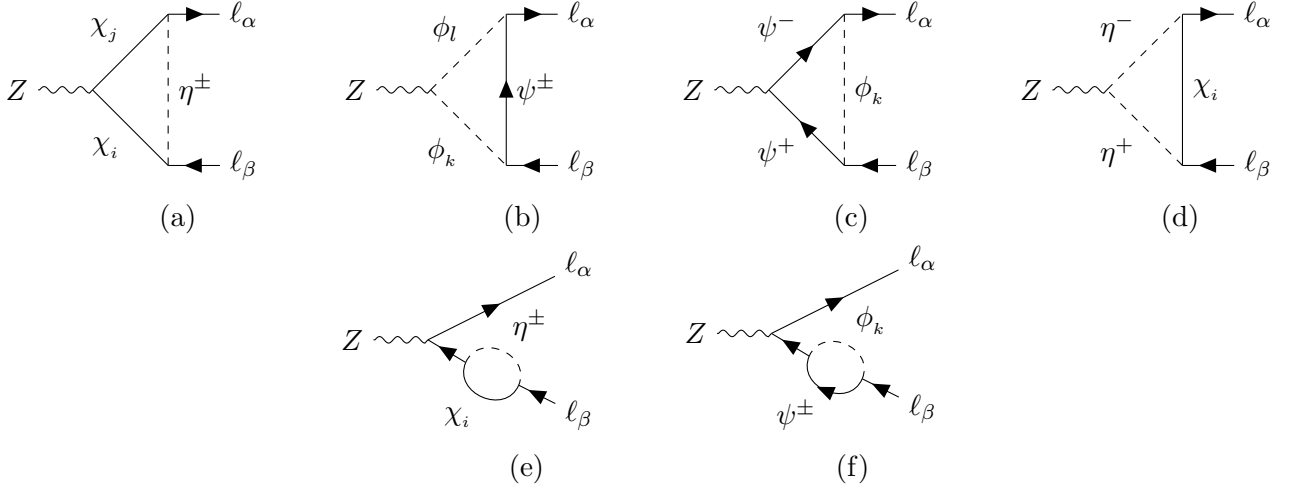

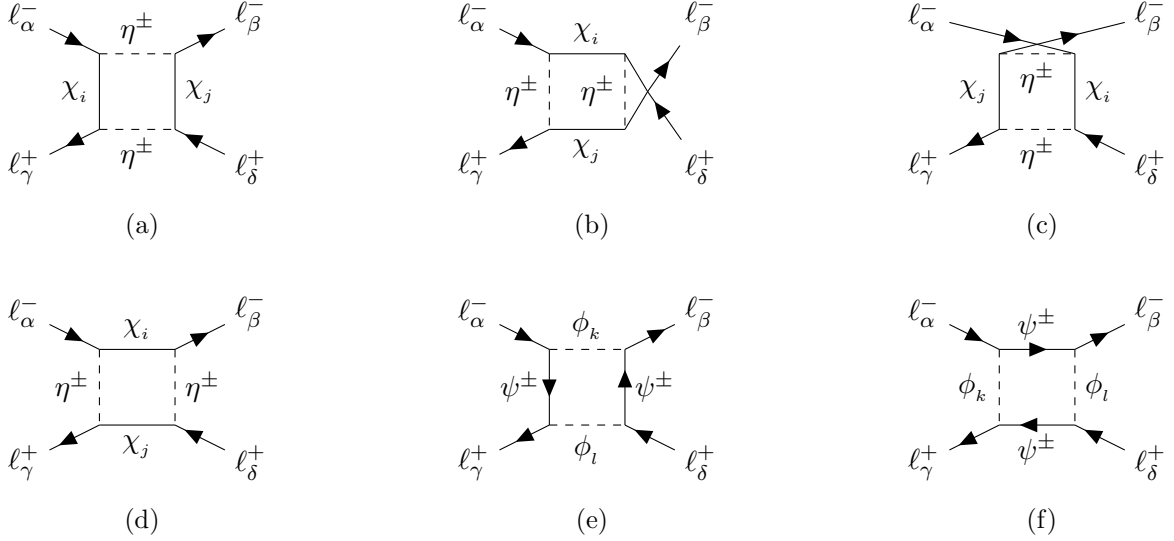
\begin{figure}[h!]
    \centering
    \begin{subfigure}[b]{0.3\textwidth}
    \centering
 \raisebox{-5mm}{    \begin{tikzpicture}
    \begin{feynman}
    \vertex (a) at (0,-0.5) {\(\ell_\gamma^+\)};
    \vertex (b) at (1,0);
    \vertex (c) at (2,0);
    \vertex (d) at (3,-0.5) {\(\ell_\delta^+\)};
    \vertex (a1) at (0,1.5) {\(\ell_\alpha^-\)};
    \vertex (b1) at (1,1);
    \vertex (c1) at (2,1);
    \vertex (d1) at (3,1.5) {\(\ell_\beta^-\)};
    \diagram* {
    (a1) -- [fermion] (b1)-- [edge label'=\( \chi_{_i}\)] (b)-- [fermion] (a),
    (d1) -- [anti fermion] (c1)-- [edge label=\( \chi_{_j}\)] (c)-- [anti fermion] (d),
    (b1) -- [scalar, edge label=\( \eta^\pm\)] (c1),
    (b) -- [scalar, edge label'=\( \eta^\pm\)] (c),
    };
    \end{feynman}
    \end{tikzpicture}
    }
            \label{}
            \caption*{(a)}
    \end{subfigure} 
    \hfill
    \begin{subfigure}[b]{0.3\textwidth}
    \centering
 \raisebox{-5mm}{    \begin{tikzpicture}
    \begin{feynman}
    \vertex (a) at (0,-0.5) {\(\ell_\gamma^+\)};
    \vertex (b) at (1,0);
    \vertex (c) at (2,0);
    \vertex (d) at (3,-0.5) {\(\ell_\delta^+\)};
    \vertex (dt) at (2.3,0.5);
    \vertex (a1) at (0,1.5) {\(\ell_\alpha^-\)};
    \vertex (b1) at (1,1);
    \vertex (c1) at (2,1);
    \vertex (d1) at (3,1.5) {\(\ell_\beta^-\)};
    \diagram* {
    (a1) -- [fermion] (b1)-- [edge label=\( \chi_{_i}\)] (c1) -- []  (dt) -- [anti fermion]  (d),
    (a) -- [anti fermion] (b)-- [edge label'=\( \chi_{_j}\)] (c) -- []  (dt)-- [fermion] (d1),
    (b1) -- [scalar, edge label'=\( \eta^\pm\)] (b),
    (c1) -- [scalar, edge label'=\( \eta^\pm\)] (c),
    };
    \end{feynman}
    \end{tikzpicture}
    }
            \label{}
            \caption*{(b)}
    \end{subfigure}
    \hfill
     \begin{subfigure}[b]{0.3\textwidth}
    \centering
 \raisebox{-5mm}{    \begin{tikzpicture}
    \begin{feynman}
    \vertex (a) at (0,-0.5) {\(\ell_\gamma^+\)};
    \vertex (b) at (1,0);
    \vertex (c) at (2,0);
    \vertex (d) at (3,-0.5) {\(\ell_\delta^+\)};
    \vertex (a1) at (0,1.5) {\(\ell_\alpha^-\)};
    \vertex (b1) at (1,1);
    \vertex (c1) at (2,1);
    \vertex (d1) at (3,1.5) {\(\ell_\beta^-\)};
    \diagram* {
    (d1) -- [anti fermion] (b1)-- [edge label'=\( \chi_{_j}\)] (b)-- [fermion] (a),
    (a1) -- [fermion] (c1)-- [edge label=\( \chi_{_i}\)] (c)-- [anti fermion] (d),
    (b1) -- [scalar, edge label'=\( \eta^\pm\)] (c1),
    (b) -- [scalar, edge label'=\( \eta^\pm\)] (c),
    };
    \end{feynman}
    \end{tikzpicture}
    }
            \label{}
            \caption*{(c)}
    \end{subfigure}
    \\
    \vspace*{5mm}
    \begin{subfigure}[b]{0.3\textwidth}
    \centering
 \raisebox{-5mm}{    \begin{tikzpicture}
    \begin{feynman}
    \vertex (a) at (0,-0.5) {\(\ell_\gamma^+\)};
    \vertex (b) at (1,0);
    \vertex (c) at (2,0);
    \vertex (d) at (3,-0.5) {\(\ell_\delta^+\)};
    \vertex (a1) at (0,1.5) {\(\ell_\alpha^-\)};
    \vertex (b1) at (1,1);
    \vertex (c1) at (2,1);
    \vertex (d1) at (3,1.5) {\(\ell_\beta^-\)};
    \diagram* {
    (a1) -- [fermion] (b1)-- [edge label=\( \chi_{_i}\)] (c1)-- [fermion] (d1),
    (a) -- [anti fermion] (b)-- [edge label'=\( \chi_{_j}\)] (c)-- [anti fermion] (d),
    (b1) -- [scalar, edge label'=\( \eta^\pm\)] (b),
    (c1) -- [scalar, edge label=\( \eta^\pm\)] (c),
    };
    \end{feynman}
    \end{tikzpicture}
    }
            \label{}
            \caption*{(d)}
    \end{subfigure}
    \hfill
    \begin{subfigure}[b]{0.3\textwidth}
    \centering
 \raisebox{-5mm}{    \begin{tikzpicture}
    \begin{feynman}
    \vertex (a) at (0,-0.5) {\(\ell_\gamma^+\)};
    \vertex (b) at (1,0);
    \vertex (c) at (2,0);
    \vertex (d) at (3,-0.5) {\(\ell_\delta^+\)};
    \vertex (a1) at (0,1.5) {\(\ell_\alpha^-\)};
    \vertex (b1) at (1,1);
    \vertex (c1) at (2,1);
    \vertex (d1) at (3,1.5) {\(\ell_\beta^-\)};
    \diagram* {
    (a1) -- [fermion] (b1)-- [fermion, edge label'=\( \psi^\pm\)] (b)-- [fermion] (a),
    (d1) -- [anti fermion] (c1)-- [anti fermion, edge label=\( \psi^\pm\)] (c)-- [anti fermion] (d),
    (b1) -- [scalar, edge label=\( \phi_{_k}\)] (c1),
    (b) -- [scalar, edge label'=\( \phi_{_l}\)] (c),
    };
    \end{feynman}
    \end{tikzpicture}
    }
            \label{}
            \caption*{(e)}
    \end{subfigure} 
    \hfill
    \begin{subfigure}[b]{0.3\textwidth}
    \centering
 \raisebox{-5mm}{    \begin{tikzpicture}
    \begin{feynman}
    \vertex (a) at (0,-0.5) {\(\ell_\gamma^+\)};
    \vertex (b) at (1,0);
    \vertex (c) at (2,0);
    \vertex (d) at (3,-0.5) {\(\ell_\delta^+\)};
    \vertex (a1) at (0,1.5) {\(\ell_\alpha^-\)};
    \vertex (b1) at (1,1);
    \vertex (c1) at (2,1);
    \vertex (d1) at (3,1.5) {\(\ell_\beta^-\)};
    \diagram* {
    (a1) -- [fermion] (b1)-- [fermion, edge label=\( \psi^\pm\)] (c1)-- [fermion] (d1),
    (a) -- [anti fermion] (b)-- [anti fermion, edge label'=\( \psi^\pm\)] (c)-- [anti fermion] (d),
    (b1) -- [scalar, edge label'=\( \phi_{_k}\)] (b),
    (c1) -- [scalar, edge label=\( \phi_{_l}\)] (c),
    };
    \end{feynman}
    \end{tikzpicture}
    }
            \label{}
            \caption*{(f)}
    \end{subfigure}
    \caption{Box diagrams contributing to $\ell_\alpha^+ \to \ell_\beta^+ \ell_\gamma^+ \ell_\delta^-$, with $i,j=1-4$ and $k,l=1-3$.}
    \label{fig:scoto:boxes}
\end{figure}

The presence of new lepton flavour violating interactions gives rise to abundant contributions to leptonic cLFV transitions and decays\footnote{In the present study we do not include a discussion of $Z$ and Higgs cLFV decays; we refer the reader to~\cite{Darricau:2025vcs}.}. 
The radiative decays emerge from the dipole interactions depicted in Fig.~\ref{fig:scoto:radiative:AMM:cLFV}, and contribute to penguin exchanges (anapole and dipole). The cLFV $Z$ interactions, summarised in Fig.~\ref{fig:scoto:Zlldiag}, also contribute to cLFV transitions, with the vector mediators attached to a charged lepton line (two same-flavoured fermions in the final state). Finally, leptonic cLFV processes also receive contributions from box diagrams, summarised in Fig.~\ref{fig:scoto:boxes}. Although included in the numerical computation, Higgs (scalar) penguins lead to subdominant contributions, and will not be discussed here.  

The complete expressions for the leptonic cLFV transitions and decays are detailed in Appendix~\ref{app:cLFV-details}, with the exception of the three-body cLFV decays, which will be discussed in the coming subsection. In Table~\ref{tab:cLFV_current_future_sensitivity} we summarise  the current experimental bounds as well as the projected experimental sensitivities for the considered cLFV observables.

\renewcommand{\arraystretch}{1.3}
\begin{table}[h!]
    \centering
    \hspace*{-2mm}{\small\begin{tabular}{|c|c|c|}
    \hline
    Observable & Current bound & Future sensitivity  \\
    \hline\hline
    $\text{BR}(\mu\to e \gamma)$    &
    \quad $<1.5\times 10^{-13}$ \quad (MEG II~\cite{MEGII:2025gzr})   &
    \quad $6\times 10^{-14}$ \quad (MEG II~\cite{Baldini:2018nnn}) \\
    $\text{BR}(\tau \to e \gamma)$  &
    \quad $<3.3\times 10^{-8}$ \quad (BaBar~\cite{Aubert:2009ag})    &
    \quad $3\times10^{-9}$ \quad (Belle II~\cite{Kou:2018nap})      \\
    $\text{BR}(\tau \to \mu \gamma)$    &
     \quad $ <4.2\times 10^{-8}$ \quad (Belle~\cite{Belle:2021ysv})  &
    \quad $10^{-9}$ \quad (Belle II~\cite{Kou:2018nap})     \\
    \hline
        $\text{BR}(\mu \to 3 e)$    &
     \quad $<1.0\times 10^{-12}$ \quad (SINDRUM~\cite{Bellgardt:1987du})    &
     \quad $10^{-15(-16)}$ \quad (Mu3e~\cite{Blondel:2013ia})   \\
    $\text{BR}(\tau \to 3 e)$   &
    \quad $<2.7\times 10^{-8}$ \quad (Belle~\cite{Hayasaka:2010np})&
    \quad $5\times10^{-10}$ \quad (Belle II~\cite{Kou:2018nap})     \\
    $\text{BR}(\tau \to 3 \mu )$    &
    \quad $<1.9\times 10^{-8}$ \quad (Belle II~\cite{Belle-II:2024sce})  &
    \quad $5\times10^{-10}$ \quad (Belle II~\cite{Kou:2018nap})     \\
    & & \quad $1.4 \times 10^{-10}$ \quad (STCF~\cite{Achasov:2023gey})\\
    & & \quad$5\times 10^{-11}$\quad (FCC-ee~\cite{Abada:2019lih})\\
     $\text{BR}(\tau^+ \to e^+ \mu^+ \mu^- )$    &
    \quad $<2.7\times 10^{-8}$ \quad (Belle~\cite{Hayasaka:2010np})  &
    \quad $5\times10^{-10}$ \quad (Belle II~\cite{Kou:2018nap})     \\
    $\text{BR}(\tau^+ \to \mu^+ e^+ e^- )$    &
    \quad $<1.8\times 10^{-8}$ \quad (Belle~\cite{Hayasaka:2010np})  &
    \quad $5\times10^{-10}$ \quad (Belle II~\cite{Kou:2018nap})     \\
    $\text{BR}(\tau^+ \to e^+ e^+ \mu^- )$    &
    \quad $<1.5\times 10^{-8}$ \quad (Belle~\cite{Hayasaka:2010np})  &
    \quad $3\times10^{-10}$ \quad (Belle II~\cite{Kou:2018nap})     \\
    $\text{BR}(\tau^+ \to \mu^+ \mu^+ e^- )$    &
    \quad $<1.7\times 10^{-8}$ \quad (Belle~\cite{Hayasaka:2010np})  &
    \quad $4\times10^{-10}$ \quad (Belle II~\cite{Kou:2018nap})     \\
    \hline
    $\text{CR}(\mu- e, \text{N})$ &
     \quad $<7 \times 10^{-13}$ \quad  (Au, SINDRUM~\cite{Bertl:2006up}) &
    \quad $10^{-14}$  \quad (SiC, DeeMe~\cite{Nguyen:2015vkk})    \\
    & &  \quad $2.6\times 10^{-17}$  \quad (Al, COMET~\cite{Krikler:2015msn,COMET:2018auw,Moritsu:2022lem})  \\
    & &  \quad $8 \times 10^{-17}$  \quad (Al, Mu2e~\cite{Bartoszek:2014mya}) \\
    \hline
    \end{tabular}}
    \caption{Current experimental bounds and future sensitivities for relevant leptonic cLFV observables. Notice that limits are given at $90\%\:\mathrm{C.L.}$, and that Belle II projected sensitivities rely on an integrated luminosity of $50\:\mathrm{ab}^{-1}$.}
    \label{tab:cLFV_current_future_sensitivity}
\end{table}
\renewcommand{\arraystretch}{1.}

\subsection{Asymmetries in 3-body cLFV decays}

As mentioned in the Introduction, in the present study we consider all generic 3-body leptonic decays, $\ell_\alpha^+ \to \ell_\beta^+ \ell_\gamma^+ \ell_\delta^-$.
For each, we investigated the associated asymmetries, which include 
$P$-, $P^\prime$- and $T$-asymmetries, as well as forward-backward asymmetries.

The 3-body decay kinematics can be conveniently cast as\footnote{For a schematic depiction of the decay, see~\cite{Darricau:2025rmu}.}
\begin{equation}\label{eq:genericdecay}
    \ell_\alpha^+ (p) \to 
    \ell_\beta^+ (k_1)\, \ell_\gamma^+ (k_2)\, \ell_\delta^-(k_3)\,.
\end{equation}
Following\cite{Bolton:2022lrg}, and working in  
the rest frame of the decaying lepton, the angle $\theta_\varepsilon$ is defined as to lie between the polarisation of the decaying lepton and the outgoing negatively charged lepton; $\phi_\varepsilon$ is the azimuthal angle in the decay plane. The conventions for the 4-momenta can be thus cast as
\begin{align}\label{eq:kinematics:convention}
    p^\mu & = (m_\alpha,0,0,0)\,,\nonumber\\ 
    k_1 & = (E_1, k_1^x, 0, k_1^z)\,,\nonumber\\
    k_2 & = (E_2, k_2^x, 0, k_2^z)\,,\nonumber\\
    k_3 & = (E_3, 0, 0, \vert \vec{k_3} \vert)\,,\nonumber\\
    \varepsilon^\mu & = (0,P \sin \theta_\varepsilon \cos \phi_\varepsilon, P \sin \theta_\varepsilon \sin \phi_\varepsilon, P \cos \theta_\varepsilon)\,, 
\end{align}
in which $(0, \vec{P})$ denotes the polarisation four-vector of the decaying lepton, with $P \equiv \vert \vec{P} \vert$, and  
\begin{equation}\label{eq:reducedmomenta}
    s_i = (p - k_i)^2\,, \quad \text{with}\quad
    s_1 + s_2 + s_3  = m_\alpha^2 + m_\beta^2 + m_\gamma^2 + m_\delta^2\,.
\end{equation}
As extensively discussed in the literature~\cite{Bolton:2022lrg, Darricau:2025rmu, Goto:2010sn}, the asymmetries can be cast as follows (starting 
from the double-differential decay width, explicitly decomposed in its spin-independent and spin-dependent parts)
\begin{align}\label{eq:DoubleDiffDec}
    \frac{d \Gamma_{\ell_\alpha^+ \to \ell_\gamma^+ \ell_\beta^+ \ell_\beta^-}}{d \Omega_\epsilon} &= \frac{\Gamma_{\ell_\alpha^+ \to \ell_\gamma^+ \ell_\beta^+ \ell_\beta^-}}{4 \pi} \left[1 + P\left(\mathcal{A}^P_{\ell_\alpha^+ \to \ell_\gamma^+ \ell_\beta^+ \ell_\beta^-}  \cos \theta_\varepsilon \right.\right. \nonumber \\
    &+ \left. \left.\mathcal{A}^{P^\prime}_{\ell_\alpha^+ \to \ell_\gamma^+ \ell_\beta^+ \ell_\beta^-}  \sin \theta_\varepsilon \cos \phi_\varepsilon + \mathcal{A}^T_{\ell_\alpha^+ \to \ell_\gamma^+ \ell_\beta^+ \ell_\beta^-}  \sin \theta_\varepsilon \sin \phi_\varepsilon\right) \right]\,.
\end{align}
The amplitudes are thus given by
\begin{align}\label{eq:asym:def}
    P\,\mathcal{A}_{\ell_\alpha^+ \to \ell_\beta^+ \ell_\gamma^+ \ell_\delta^-}^P & = 
    \frac{1}{\Gamma_{\ell_\alpha^+ \to \ell_\beta^+ \ell_\gamma^+ \ell_\delta^-}} 
    \,\int_0^{2 \pi} d \phi_\varepsilon \left(\int_{0}^1 - \int_{-1}^0 \right) d \cos \theta_\varepsilon \int_\Omega d \Omega\, \frac{d \Gamma_{\ell_\alpha^+ \to \ell_\beta^+ \ell_\gamma^+ \ell_\delta^-}}{d \Omega_\varepsilon \,d \Omega}\,,\nonumber\\
    P\,\mathcal{A}_{\ell_\alpha^+ \to \ell_\beta^+ \ell_\gamma^+ \ell_\delta^-}^{P^\prime} & = \frac{1}{\Gamma_{\ell_\alpha^+ \to \ell_\beta^+ \ell_\gamma^+ \ell_\delta^-}} 
    \,  \left(\int_{-\frac{\pi}{2}}^\frac{\pi}{2} - \int_{\frac{\pi}{2}}^\frac{3 \pi}{2} \right) d \phi_\varepsilon \int_{-1}^1 d \cos \theta_\varepsilon \int_\Omega \,d \Omega \frac{d \Gamma_{\ell_\alpha^+ \to \ell_\beta^+ \ell_\gamma^+ \ell_\delta^-}}{d \Omega_\varepsilon \,d \Omega}\,,\nonumber\\
    P\,\mathcal{A}_{\ell_\alpha^+ \to \ell_\beta^+ \ell_\gamma^+ \ell_\delta^-}^T & = \frac{1}{\Gamma_{\ell_\alpha^+ \to \ell_\beta^+ \ell_\gamma^+ \ell_\delta^-}} 
    \,  \left(\int_{0}^\pi - \int_{- \pi}^0 \right) d \phi_\varepsilon \int_{-1}^1 d \cos \theta_\varepsilon \int_\Omega \,d \Omega \frac{d \Gamma_{\ell_\alpha^+ \to \ell_\beta^+ \ell_\gamma^+ \ell_\delta^-}}{d \Omega_\varepsilon \,d \Omega}\,.
\end{align}
Moreover one trivially has for the total cLFV 3-body decay width
\begin{equation}
        \Gamma_{\ell_\alpha^+ \to \ell_\beta^+ \ell_\gamma^+ \ell_\delta^-}  =  \int_0^{2 \pi} d \phi_\varepsilon \int_{-1}^1 d \cos \theta_\varepsilon \int_\Omega d \Omega \,\frac{d \Gamma_{\ell_\alpha^+ \to \ell_\beta^+ \ell_\gamma^+ \ell_\delta^-}}{d \Omega_\varepsilon \,d \Omega}\,.
\end{equation}

As expected, a number of distinct operators can contribute to the 3-body decay widths, including dipole, $Z$ and Higgs penguins, as well as box diagrams. 
The full amplitudes for the transitions have been computed for the ``T1-2-A'' scotogenic realisation~\cite{Darricau:2025vcs} using the effective operator basis of~\cite{Abada:2014kba}, and below we present the analytic expressions, considering the $\ell_\alpha \to 3 \ell_\beta$ decays for simplicity (and not the generic and more involved $\ell_\alpha^+ \to \ell_\beta^+ \ell_\gamma^+ \ell_\delta^-$ transitions):
\begin{align}
    \label{eqn:analytic_lto3lp}
        \text{BR}\left( \ell_{\alpha} \rightarrow 3\ell_{\beta} \right) &= \frac{m_{\ell_{\alpha}}^{5}}{512\pi^{3} \Gamma_{{\ell_{\alpha}}}} \left[ e^{4} \left( \lvert K_{2}^{L} \rvert^{2} + \lvert K_{2}^{R} \rvert^{2} \right) \left( \frac{16}{3} \ln \frac{m_{\ell_{\alpha}}}{m_{\ell_{\beta}}} - \frac{22}{3} \right) \right. \nonumber \\
        &+ \frac{1}{24} \left( \lvert A_{LL}^{S} \rvert^{2} + \lvert A_{RR}^{S} \rvert^{2} \right) + \frac{1}{12} \left( \lvert A_{LR}^{S} \rvert^{2} + \lvert A_{RL}^{S} \rvert^{2} \right)\nonumber  \\
        &+ \frac{2}{3} \left( \lvert \hat{A}_{LL}^{V} \rvert^{2} + \lvert \hat{A}_{RR}^{V} \rvert^{2} \right) + \frac{1}{3} \left( \lvert \hat{A}_{LR}^{V} \rvert^{2} + \lvert \hat{A}_{RL}^{V} \rvert^{2} \right) + 6 \left( \lvert \hat{A}_{LL}^{T} \rvert^{2} + \lvert \hat{A}_{RR}^{T} \rvert^{2} \right) \nonumber \\
        &+ \frac{e^{2}}{3} \left( K_{2}^{L} \,A_{RL}^{S*} + K_{2}^{R} A_{LR}^{S*} + \text{c.c} \right) - \frac{2e^{2}}{3} \left( K_{2}^{L}\, \hat{A}_{RL}^{V*} + K_{2}^{R} \,\hat{A}_{LR}^{V*} + \text{c.c} \right) \nonumber \\
        &- \frac{4e^{2}}{3} \left( K_{2}^{L} \,\hat{A}_{RR}^{V*} + K_{2}^{R}\, A_{LL}^{V*} + \text{c.c} \right) \nonumber \\
        &\left. -\frac{1}{2} \left( A_{LL}^{S} \,A_{LL}^{T*} + A_{RR}^{S}\, A_{RR}^{T*} + \text{c.c} \right) -\frac{1}{6} \left( A_{LR}^{S} \,\hat{A}_{LR}^{V*} + A_{RL}^{S} \,\hat{A}_{RL}^{V*} + \text{c.c} \right) \right]\,,
\end{align}
in which
\begin{equation}
  \hat{A}_{XY}^{V} = A_{XY}^{V} + e^{2} K_{1}^{X}\,, \quad \text{with }\quad X, Y = L, R\,. 
\end{equation}
In the above, $A$ (and $\hat{A}$) correspond to the 4-fermion ($4 \ell$) form factors, which can be decomposed into photon, $Z$ and Higgs penguin as well as box contributions.
Let us mention that the anapole (dipole) $\gamma$-penguin 
contributions are referred to as $K_{1}^{L/R}$ ($K_{2}^{L/R}$) where $K_{2}^{R} = 2 c_{R}/m_{\ell_{_\alpha}}$ and $L \leftrightarrow R $. 
The detailed analytical expressions of the remaining operators can be found in Appendix~\ref{app:cLFV-details}.

Similarly to the branching ratio expression, one can derive simple expressions for the asymmetries
\begin{align}
    \label{eqn:analytic_Asymlto3lp}
        \mathcal{A}^P_{\ell_{\alpha} \rightarrow 3\ell_{\beta}} & \approx \frac{m_{\ell_{\alpha}}^{5}}{36\ 864\pi^{3} \Gamma_{\ell_{\alpha} \rightarrow 3\ell_{\beta}}} \left[ 48 e^{4} \left( \lvert K_{2}^{R} \rvert^{2} - \lvert K_{2}^{L} \rvert^{2} \right) \left( 8 \ln \frac{m_{\ell_{\alpha}}}{m_{\ell_{\beta}}} - 23 \right) \right. \nonumber \\
        &+ 3 \left( \lvert A_{RR}^{S} \rvert^{2} - \lvert A_{LL}^{S} \rvert^{2} \right) - 2 \left( \lvert A_{RL}^{S} \rvert^{2} - \lvert A_{LR}^{S} \rvert^{2} \right)\nonumber  \\
        &+ 48 \left( \lvert \hat{A}_{RR}^{V} \rvert^{2} - \lvert \hat{A}_{LL}^{V} \rvert^{2} \right) - 8 \left( \lvert \hat{A}_{RL}^{V} \rvert^{2} - \lvert \hat{A}_{LR}^{V} \rvert^{2} \right) + 432 \left( \lvert \hat{A}_{RR}^{T} \rvert^{2} - \lvert \hat{A}_{LL}^{T} \rvert^{2} \right) \nonumber \\
        &+ 24 e^{2} \left( K_{2}^{R} \,A_{LR}^{S*} - K_{2}^{L} A_{RL}^{S*} + \text{c.c} \right) - 48 e^{2} \left( K_{2}^{R}\, \hat{A}_{LR}^{V*} - K_{2}^{L} \,\hat{A}_{RL}^{V*} + \text{c.c} \right) \nonumber \\
        &+ 96 e^{2} \left( K_{2}^{R} \,\hat{A}_{LL}^{V*} - K_{2}^{L}\, A_{RR}^{V*} + \text{c.c} \right) \nonumber \\
        &\left. - 36 \left( A_{RR}^{S} \,A_{RR}^{T*} - A_{LL}^{S}\, A_{LL}^{T*} + \text{c.c} \right) + 4 \left( A_{RL}^{S} \,\hat{A}_{RL}^{V*} - A_{LR}^{S} \,\hat{A}_{LR}^{V*} + \text{c.c} \right) \right]\,,\\
        \mathcal{A}^{P^\prime}_{\ell_{\alpha} \rightarrow 3\ell_{\beta}} & \approx \frac{m_{\ell_{\alpha}}^{5}}{40\ 320\pi^{3} \Gamma_{\ell_{\alpha} \rightarrow 3\ell_{\beta}}} \left[ 252 e^{4} \left( \lvert K_{2}^{R} \rvert^{2} - \lvert K_{2}^{L} \rvert^{2} \right) + \left( \lvert A_{RL}^{S} \rvert^{2} - \lvert A_{LR}^{S} \rvert^{2} \right) + 4 \left( \lvert \hat{A}_{RL}^{V} \rvert^{2} - \lvert \hat{A}_{LR}^{V} \rvert^{2} \right) \right. \nonumber \\
        & \left. - 12 e^{2} \left( K_{2}^{R} \,A_{LR}^{S*} - K_{2}^{L} A_{RL}^{S*} + \text{c.c} \right) + 24 e^{2} \left( K_{2}^{R}\, \hat{A}_{LR}^{V*} - K_{2}^{L} \,\hat{A}_{RL}^{V*} + \text{c.c} \right) \right. \nonumber \\
        & \left. - 36 e^{2} \left( K_{2}^{R} \,\hat{A}_{LL}^{V*} - K_{2}^{L}\, A_{RR}^{V*} + \text{c.c} \right) - 2 \left( A_{RL}^{S} \,\hat{A}_{RL}^{V*} - A_{LR}^{S} \,\hat{A}_{LR}^{V*} + \text{c.c} \right) \right]\,,\\
        \mathcal{A}^{T}_{\ell_{\alpha} \rightarrow 3\ell_{\beta}} & \approx \frac{e^{2} m_{\ell_{\alpha}}^{5}}{1\ 680\pi^{3} \Gamma_{\ell_{\alpha} \rightarrow 3\ell_{\beta}}} Im \! \left( K_2^L \left( 2 A_{RL}^{V*} - 3 A_{RR}^{V*} - A_{RL}^{S*}\right) + K_2^R \left( 2 A_{LR}^{V*} - 3 A_{LL}^{V*} - A_{LR}^{S*}\right)\right)\,,
\end{align}
with $\Gamma_{\ell_{\alpha} \rightarrow 3\ell_{\beta}} \equiv \Gamma_{\ell_\alpha} \text{BR}(\ell_\alpha \to 3 \ell_\beta)$. Further details on the computation of the asymmetries (and additional discussion, albeit in a different NP context) can be found in~\cite{Darricau:2025rmu}.

In what concerns the decay widths, when compared to the numerical integration over the phase space, the first order  asymptotic expansion in $m_{\ell_\beta} / m_{\ell_\alpha} \sim 0$ leads to very small numerical errors. However, this is not the case regarding the asymmetries, especially in association with  $\tau \to 3 \mu$ decays, for which the first order approximation leads to large numerical errors~\cite{Darricau:2025rmu}. In our study we will systematically rely on the full numerical integrations for the distinct observables.

\section{cLFV observables in the scotogenic ``T1-2-A''  variant}\label{sec:results}

Before discussing the expected outcome for the asymmetries in cLFV decays, in view of a change in paradigm concerning the muon anomalous magnetic moment, and especially given the much improved numerical survey of the ``T1-2-A'' parameter space, we first revisit the impact of this class of scotogenic models for cLFV observables (and potential peculiar correlations).

\subsection{Updated prospects for cLFV}
Firstly, it is important to highlight that the need for a non-negligible deviation between observation and the SM prediction for $(g-2)_\mu$ strongly favoured regions in the model's parameter space leading to important dipole contributions\footnote{As clear from Fig.~\ref{fig:scoto:radiative:AMM:cLFV}, there is a manifest common origin of the NP contributions to 
charged lepton electric and magnetic moments (i.e. $\alpha=\beta$) and the radiative cLFV decays (for $\alpha \neq \beta$), flavour-violating vertices aside.}. Even to quell a very mild tension (for example, $\Delta a_\mu \sim \mathcal{O}(1 \sigma)$), the required enhancement of the dipole contribution would lead to regimes of dipole dominance for cLFV observables (3-body decays and muon-electron conversion in nuclei); in turn, this favours a correlation between observables that has been long identified and extensively discussed in the literature~\cite{Vicente:2014wga,Rocha-Moran:2016enp,Boruah:2021qlf,Alvarez:2023dzz,Kang:2021jmi,Borah:2023hqw,Darricau:2025vcs}.
As mentioned in Section~\ref{sec:pheno_impact}, the agreement between observation and SM expectation has rendered  $(g-2)_\mu$ a new SM precision observable, leading instead to constraints on the dipole contributions. 

In view of the above discussion, one can now have competing contributions to cLFV observables (other than radiative decays) arising from very distinct operators. Moreover, the implemented DE-MCMC scheme  allows a far more efficient survey of the ``T1-2-A'' parameter space, and accessing regimes which were previously hard to explore. All this warrants a brief summary of the prospects for cLFV in the ``T1-2-A'' scotogenic realisation, which will also prove instrumental for the subsequent discussion of the asymmetries in 3-body cLFV decays.

\begin{figure}[h!]
\centering
\includegraphics[width=0.45 \textwidth] {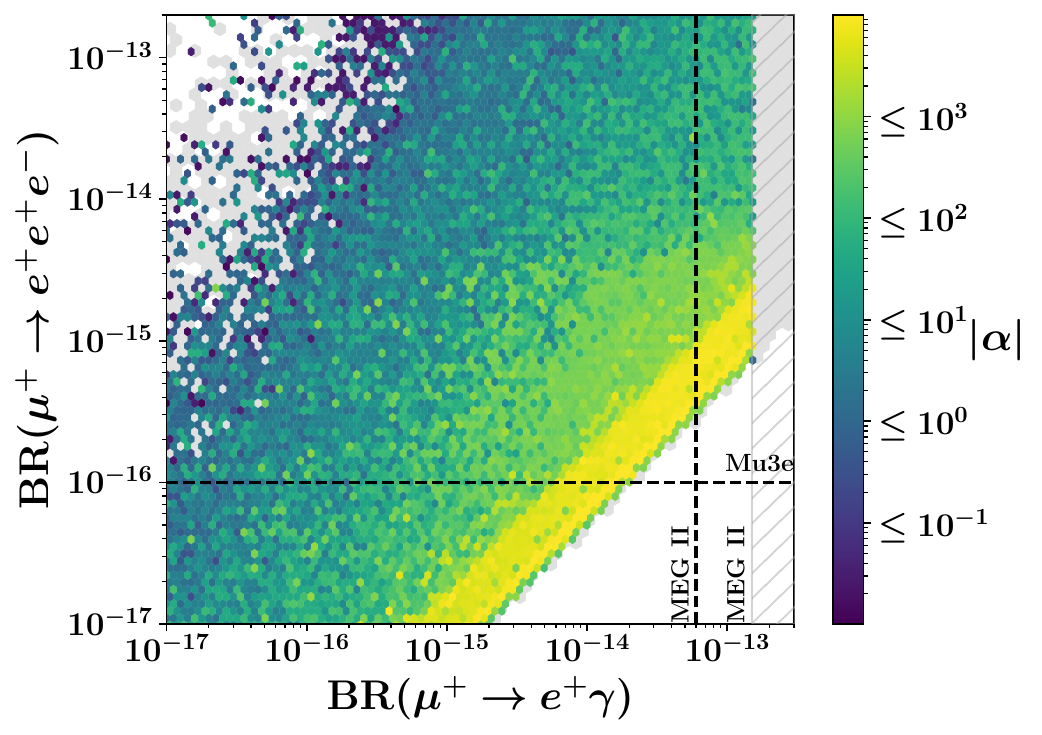}
\hspace*{8mm}
\includegraphics[width=0.45 \textwidth] {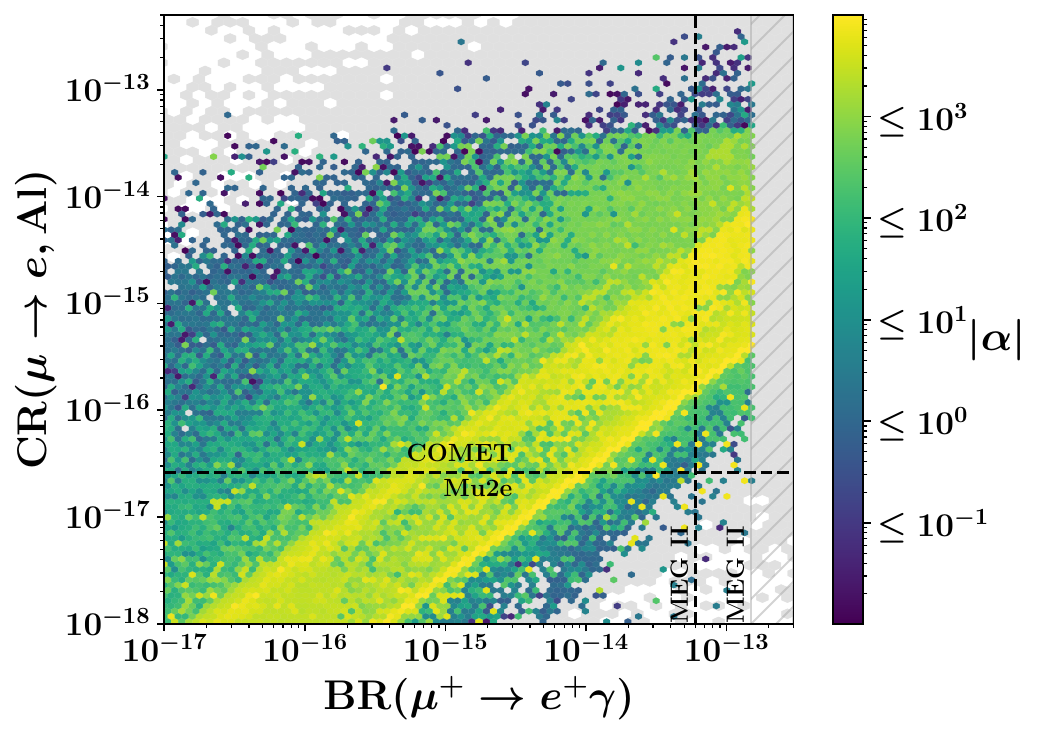}
\vspace*{3mm}
\\
\includegraphics[width=0.45 \textwidth] {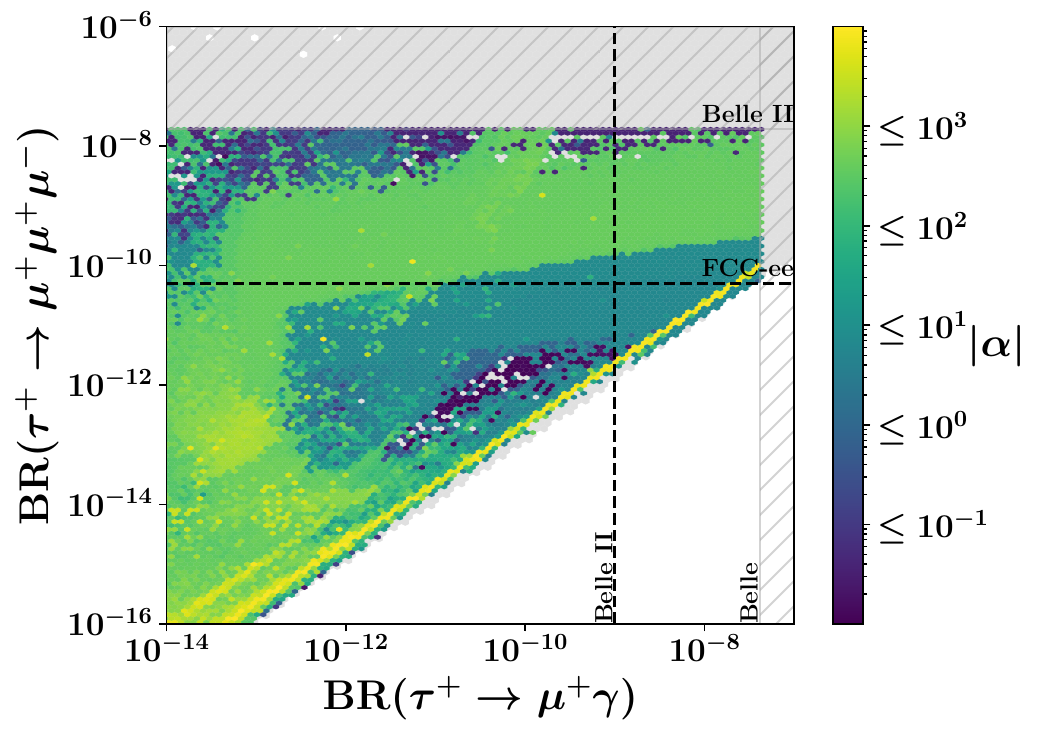}
\hspace*{8mm}
\includegraphics[width=0.45 \textwidth] {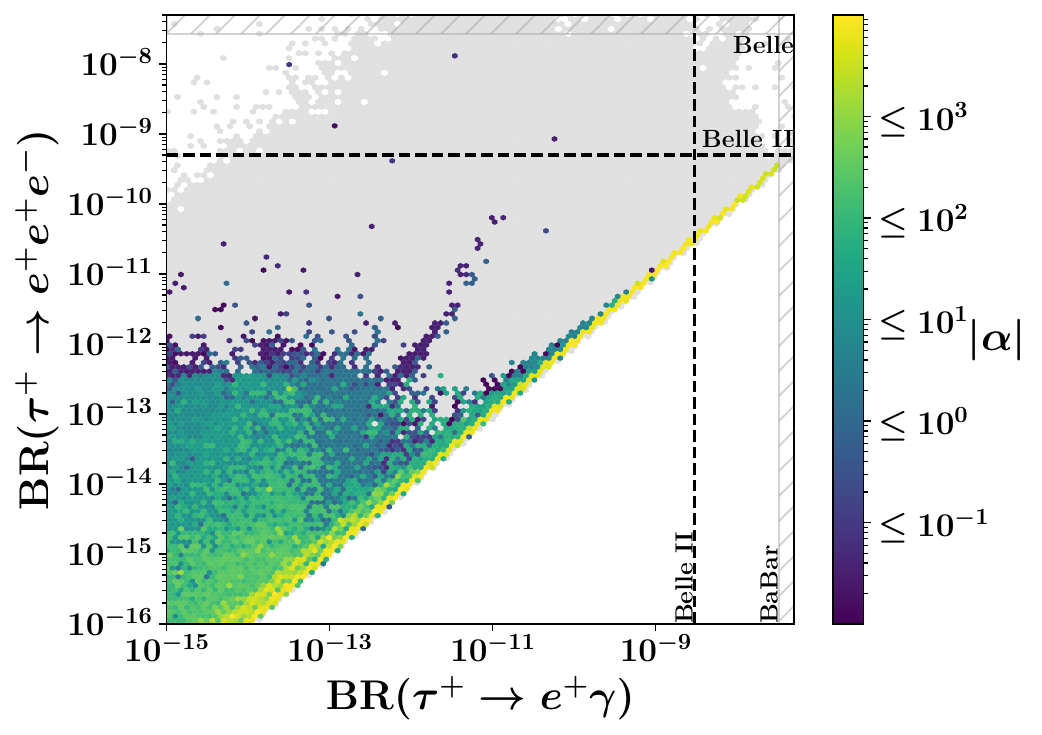}
\caption{cLFV prospects for the ``T1-2-A'' scotogenic realisation: on the upper row, $\text{BR}(\mu^+ \to e^+ e^+ e^-)$ vs. $\text{BR}(\mu^+ \to e^+ \gamma)$ (left)
and $\text{CR}(\mu \to e, \text{Al})$ vs. $\text{BR}(\mu^+ \to e^+ \gamma)$ (right); the lower panels depict 
$\text{BR}(\tau^+ \to \mu^+ \mu^+ \mu^-)$ vs. $\text{BR}(\tau^+ \to \mu^+ \gamma)$, and  $\text{BR}(\tau^+ \to e^+ e^+ e^-)$ vs. $\text{BR}(\tau^+ \to e^+ \gamma)$, respectively on the left and right panels. For all plots, the colour range denotes values for the trilinear coupling $\alpha$ (in GeV), cf. Eq.~(\ref{eq:Vscalar}), see vertical palette. Grey points are excluded due to violating at least one experimental constraint.
}
\label{fig:l3l_NPalpha}
\end{figure}

Figure~\ref{fig:l3l_NPalpha} illustrates several of the points previously mentioned. Notice that phenomenological viable points now spread over most of the considered ranges for the observables, so that any formerly existing correlation is now lost. Interestingly, for very large values of the trilinear coupling $\alpha$, in particular above the TeV, one is led to the dipole dominance regime, and accordingly, to correlated predictions for the 3-body and radiative decays, both for muon and tau leptons. Likewise, a similar scenario is observed for neutrinoless muon-electron conversion and radiative muon decays. In a sense, the yellow band corresponds to the correlated prediction for the two observables identified in previous studies~\cite{Alvarez:2023dzz, Darricau:2025vcs}. 
Despite the loss of the correlating patterns between observables, which was formerly a powerful falsifying probe of this class of models,  one verifies that there are indeed excellent prospects in what concerns observability: with the exception of the $\tau-e$ modes, the cLFV contributions can almost fully saturate the future experimental sensitivity of the different facilities. For the  
$\tau-e$ modes, there is a very marginal possibility of discovering the radiative decays at Belle II (the points within reach for $\tau \to 3e$ decays - associated with tiny values of $\alpha$ - are not statistically significant). Finally, let us notice that 
while we did numerically assess all possible cLFV $\tau$ three body decays, i.e. $\ell_\alpha^+ \to 3 \ell_\beta^+$ (same-flavour final state leptons), 
$\ell_\alpha^+ \to \ell_\beta^+ \ell_\gamma^+ \ell_\gamma^-$ (final state composed of distinguishable leptons) and $\ell_\alpha^+ \to \ell_\beta^+ \ell_\beta^+ \ell_\delta^-$ (transitions in which the lepton flavour changes by two units), we only discuss the same-flavour final state modes; this stems from having the rates associated with different flavour final states typically lying well below future experimental sensitivity. 

\begin{figure}[h!]
\centering
\includegraphics[width=0.48 \textwidth] {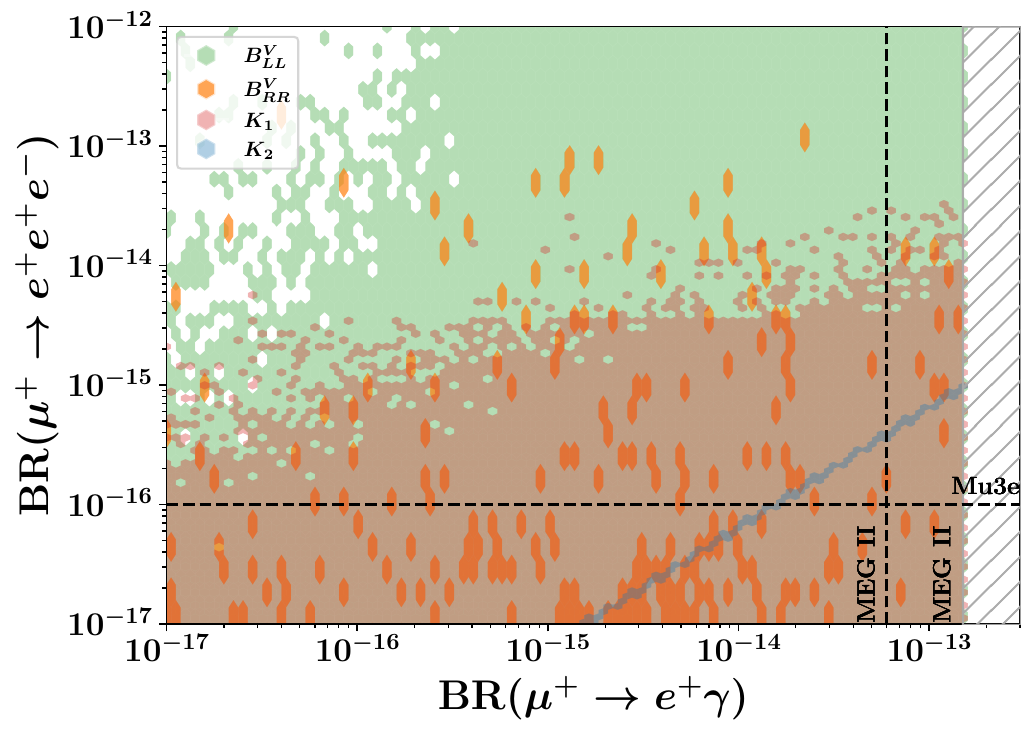}\hspace*{8mm}
\caption{Contributions of different operators to the rates for $\mu^+ \to e^+ \gamma$
and $\mu^+ \to e^+ e^- e^+$: $K_1$, $K_2$ and $B^V_{LL,RR}$, respectively depicted in maroon, blue, green and orange.
}
\label{fig:Coeff_contrib}
\end{figure}

\bigskip
For completeness, and also to prepare the discussion on the asymmetries, it is worth commenting on the nature of the  contributions to the cLFV observables. This is done in Fig.~\ref{fig:Coeff_contrib}, focusing for simplicity on $\mu-e$ transitions. We accordingly display the individual contribution of each operator to the total decay widths, obtained while forcefully setting to zero all other operators. 

Firstly and as expected, notice that the contributions from the dipole operator (i.e. proportional to $K_2^{L/R}$, see for instance Eq.~(\ref{eqn:analytic_lto3lp})) are associated with a strict correlation between the radiative and the 3-body muon decays. However, and in agreement with the findings of Fig.~\ref{fig:l3l_NPalpha}, several other operators can now lead to dominant contributions. 

It is particularly interesting to observe the significant contributions of box diagrams (corresponding to  $B^V_{LL,RR}$), and albeit to a smaller extent, those of anapole penguins (which appear scattered through the plot). 
Concerning the anapole contributions, let us recall that $K_1^R$ directly scales with the coupling bilinear $g_R^\alpha {g_R^\beta}^*$ (as seen from Eq.~(\ref{eq:anapoleFF})). 
Large contributions can thus be encountered upon the exploration of regimes characterised by sizeable values of the latter right-handed Yukawa-like coupling combination. The left-handed anapole form-factor is in general subdominant in comparison to its right-handed counterpart (also scaling with a combination of left-handed NP Yukawa-like couplings in the matrix $\mathcal{G}$). Concerning the dominant contributions to $\mu \to 3e$ arising from the left-left vector-like box form factors, we notice that the latter do prefer small values of $\alpha \lesssim 1$~TeV and, like the anapole, they also strongly depend on combinations of left-handed couplings in $\mathcal{G}$.
Further contributions from other operators are marginal, and have not been displayed.

Finally, it is important to emphasise that contrary to the study of~\cite{Alvarez:2023dzz}, 
we have explored regimes featuring wide ranges for the new Yukawa-like couplings (both those in $\mathcal{G}$,  and $g_R$ as well), in particular allowing for sizeable values for all generations. Such larges values of the couplings (while still perturbative) open a wide door to contributions clearly departing from the former ``dipole correlation'' bands.

Although we do not display them here, one encounters a similar scenario for $\tau \to 3 \mu$ decays, with significant contributions arising from a variety of operators. We briefly mention that the dominant contributions now stem from right-right vector-like box form factors, proportional to the combination of couplings $g_R^\alpha {g_R^\beta}^*$. There still are
non-negligible contributions from $B^V_{LL}$, while anapole contributions and $Z-$penguin contributions turn out to be marginal.

\subsection{Asymmetries in cLFV 3-body decays}

In view of the loss of correlation between observables, which was formerly peculiar to this class of scotogenic models,
it becomes all the more important to extend the number of cLFV observables under consideration. In what follows, we explore the prospects for several asymmetries in cLFV 3-body muon and tau decays.

\paragraph{Experimental prospects}

As muons in the Mu3e experiment~\cite{Mu3e:2020gyw} are produced from pion decays, their initial polarisation is precisely known (and close to $100\%$); since the discrimination of the signal from the background requires precise reconstruction of the momentum and common vertex of the final state electrons, in the event of a $\mu \to 3 e$ signal all conditions are in principle met to allow measuring the here discussed asymmetries (given a sufficient number of events).

The prospects for measuring cLFV tau decay asymmetries 
would be more favourable for $\tau$ production at the $Z$-pole, relying on the correlation of $\tau$-spins between the tau ``tag-hemisphere'' and the cLFV ``measurement-hemisphere''. Depending on the SM-allowed tau decay channels, one can expect ``spin-analysing powers'' ranging from $\simeq 30\%$ (in the leptonic decays), and up to $100\%$ in $\tau\to\pi\nu$~\cite{Hagiwara:1989fn,Alemany:1991ki}. As discussed in~\cite{Darricau:2025rmu}, a determination of the $\tau$-spins on an event-by-event basis would allow having highly-polarised (sub-)samples\footnote{One could envisage samples exhibiting polarisations close to $\sim 100 \%$, this being limited only by the $\tau$-spin correlation of $\sim99\%$, and by the experimental resolution of the $\pi$-momentum. }, albeit with reduced statistics.  
Similar considerations could apply to dedicated searches for tau cLFV 3-body decays at lower energies, as at Belle-II~\cite{Belle-II:2018jsg,Banerjee:2022sgf} and at the proposed super-tau-charm-factory (STCF)~\cite{Achasov:2023gey} (recalling that in the latter case polarisation of one or both initial $e^+/  e^-$ beams could further enhance $\tau$ polarisation). 

\paragraph{Expected range for the asymmetries}
Having obtained an update on the potential contributions of the present NP model to distinct cLFV observables, we now proceed to explore to which extent one can possibly use the $T$, $P$ and $P^\prime$ asymmetries\footnote{Since we will focus here on same-flavoured final states, we do not address forward-backward asymmetries.} in 3-body muon and tau decays to further probe this class of scotogenic models. 

\begin{figure}[h!]
\centering
\includegraphics[width=0.48 \textwidth] {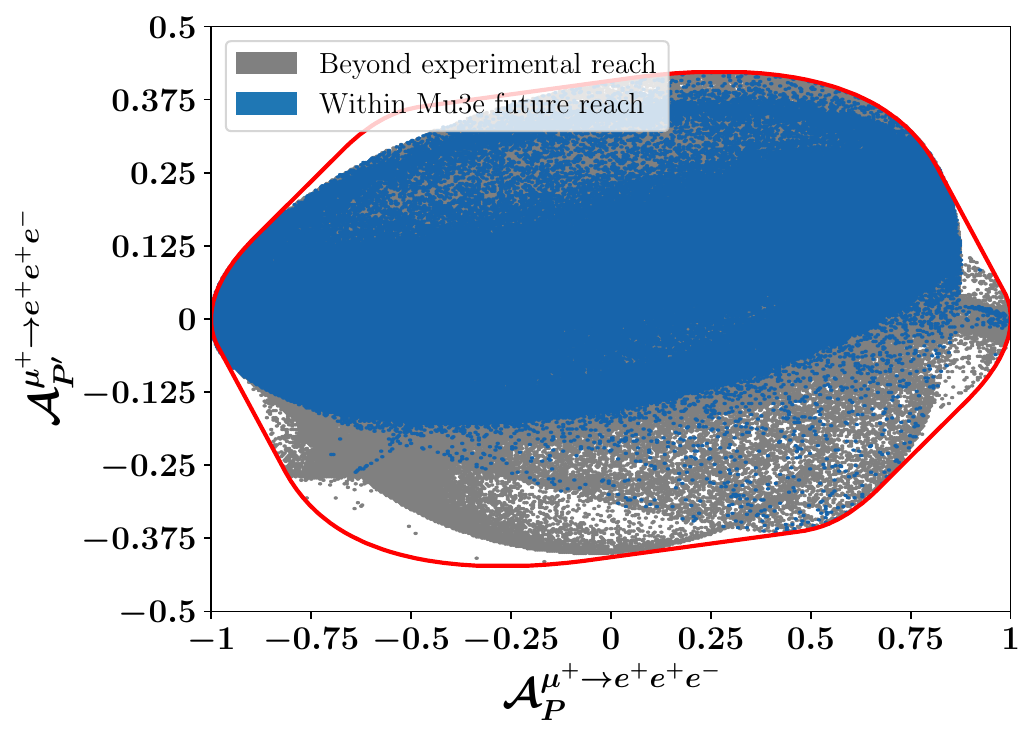}
\hspace*{3mm}
\includegraphics[width=0.48 \textwidth] {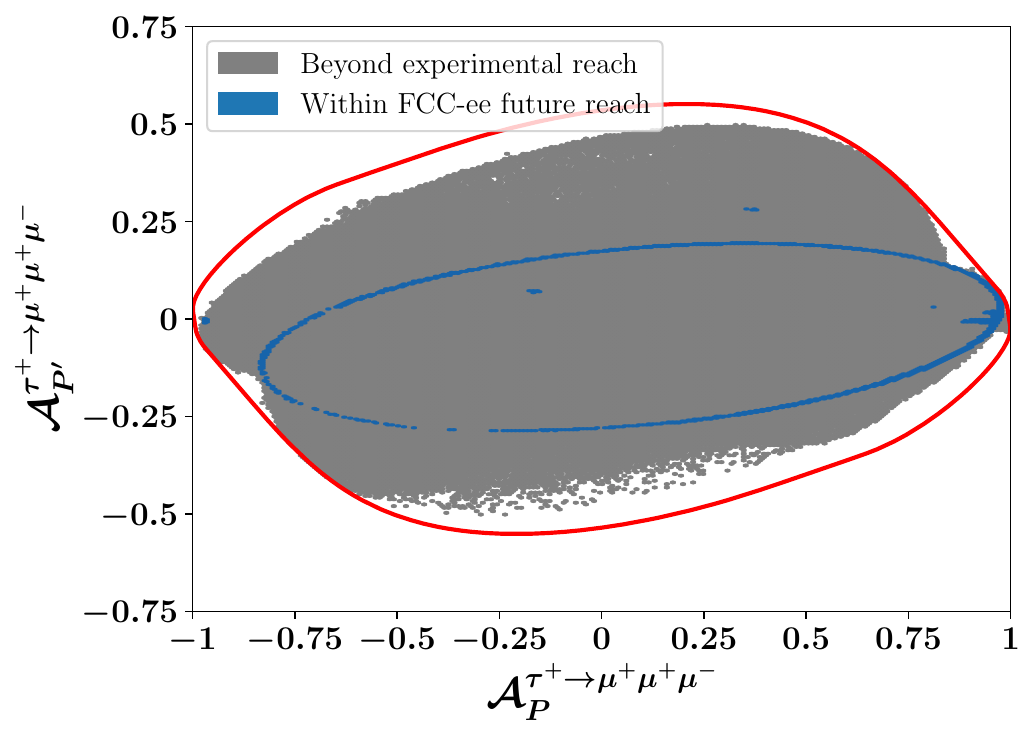}
\hspace*{3mm}
\\
\includegraphics[width=0.48 \textwidth] {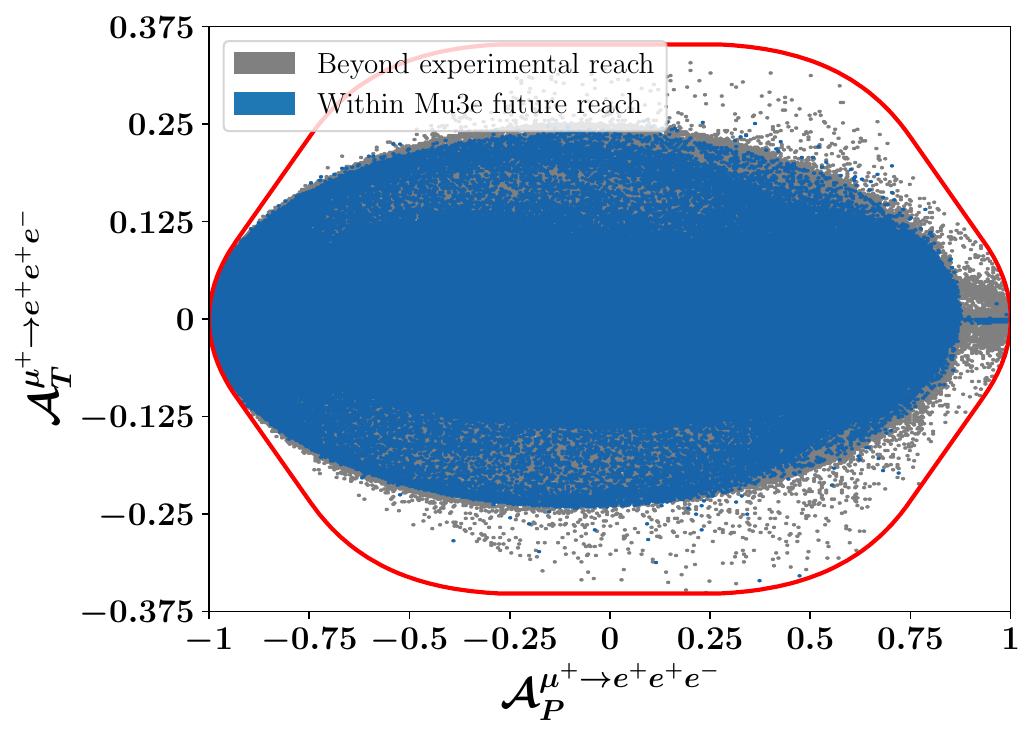}
\includegraphics[width=0.48 \textwidth] {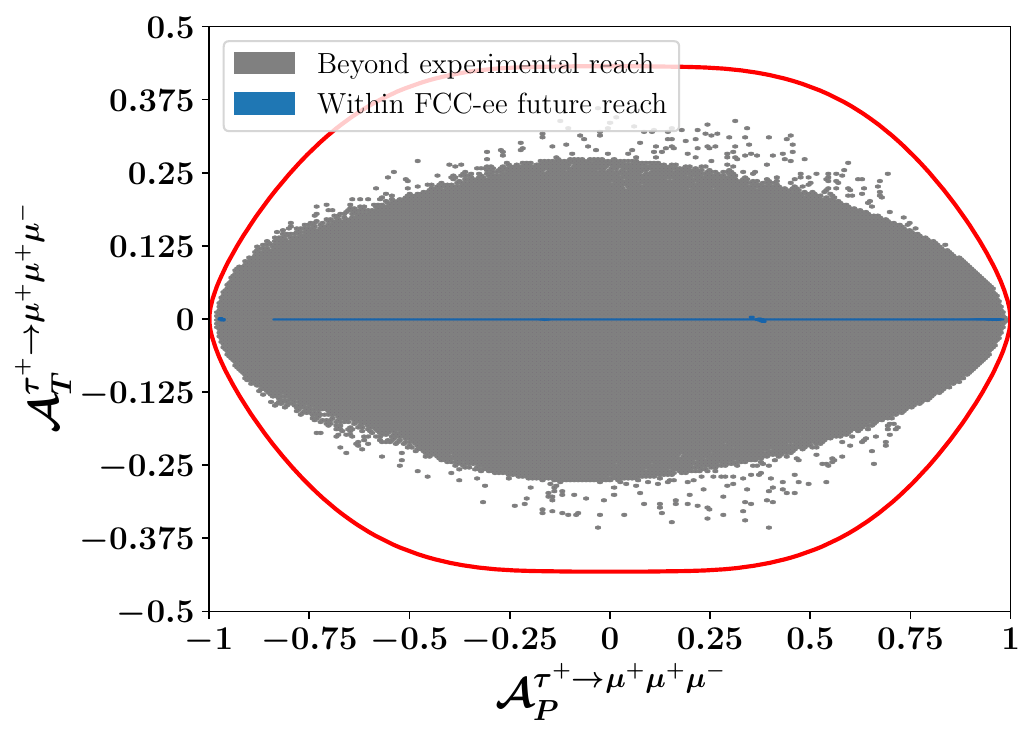}
\hspace*{3mm}
\\
\includegraphics[width=0.48 \textwidth] {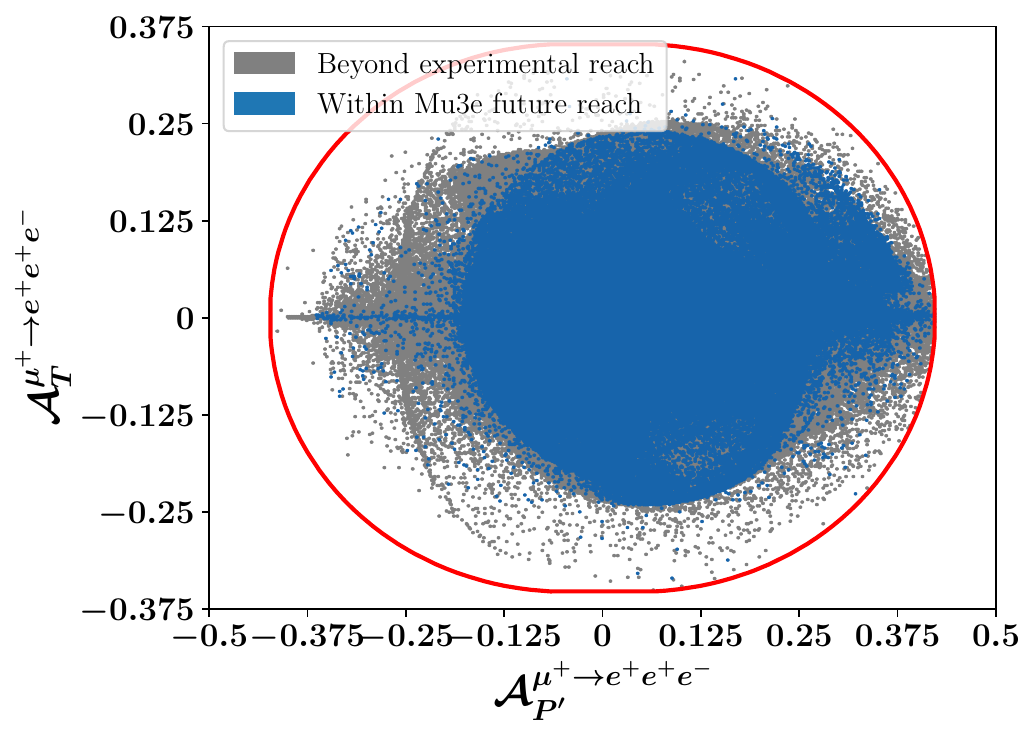}
\includegraphics[width=0.48 \textwidth] {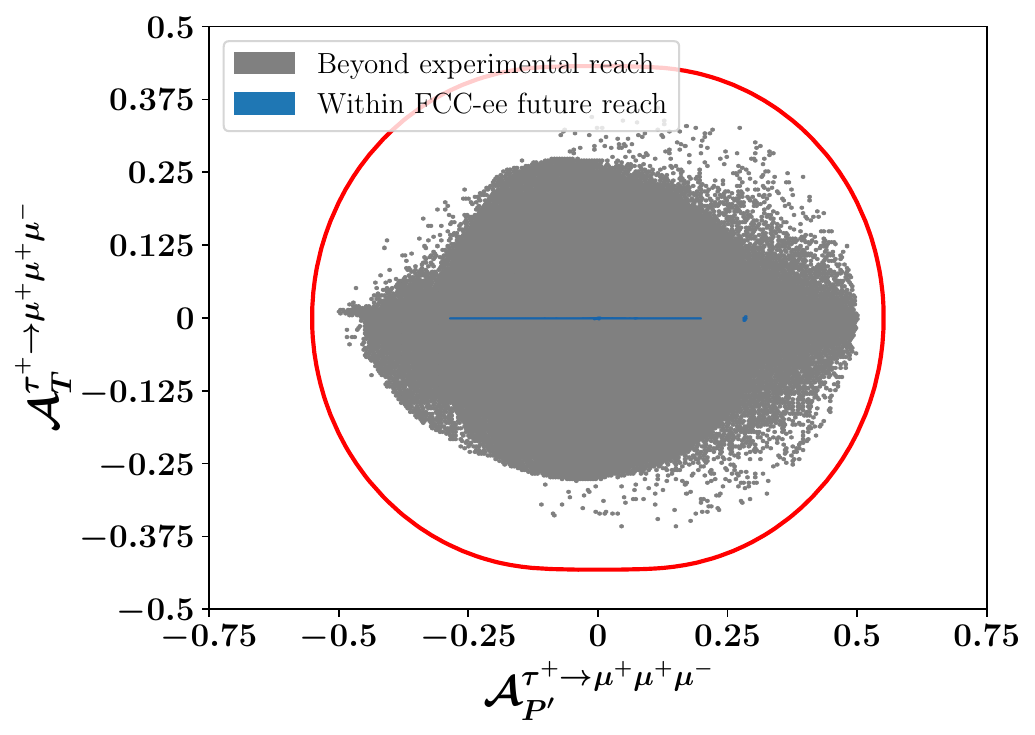}
\caption{Joint prospects for $\mu \to 3 e$ asymmetries (left panels) and for $\tau \to 3 \mu$ 
asymmetries (right panels). 
All points displayed are in agreement with current experimental bounds; the coloured (grey) points are within (beyond) future sensitivity for the decays under consideration. The red line corresponds to the maximal values for the asymmetries.}
\label{fig:l3l_Asym_vs_Asym}
\end{figure}

As a rapid and comprehensive means to illustrate our findings, in Fig.~\ref{fig:l3l_Asym_vs_Asym} we present the prospects for $\mathcal{A}_{T, P, P^\prime}$ in 
$\mu \to 3e$ and $\tau \to 3\mu$ decays (in view of the poor prospects for observation of the other cLFV $\tau$ 3-body decay modes, we only consider $\tau \to 3\mu$). All the displayed points are phenomenologically viable (that is, in agreement with all the considered constraints), the colour code only distinguishing between the prospects for future observation of the associated 3-body decays. For completeness, we also include the maximal theoretical contour for the asymmetries, which is inferred from simultaneously maximising the analytical expressions, see Eqs.~(\ref{eqn:analytic_Asymlto3lp}).

As expected, while $\mathcal{A}_P$ ranges up to $\pm 1$ (independently of the flavour of the decaying charged lepton), the absolute values of $\mathcal{A}_T$ and $\mathcal{A}_{P^\prime}$
are typically bound to lie below $50\%$. 
Concerning the asymmetries in muon decays displayed on the left column of Fig.~\ref{fig:l3l_Asym_vs_Asym}, the allowed points fill an elliptic surface, suggesting an interference between 
non-negligible contributions from at least three operators. In sharp contrast, and regarding the tau decay asymmetries (in the right column), the  elliptic band reflects the interference of exactly 2 operators\footnote{On the top right plot, a finite set of points lie outside of the ellipse band, corresponding to very specific (and statistically non-representative regimes, in which another form factor dominates).}. 

For $\mu \to 3e$ decays, the asymmetries can be quite large, up to 25\%, or even larger for $\mathcal{A}_{P,P^\prime}$. The clear potential for experimental observation of the decay rate and associated asymmetries allows strengthening the case for such a model in the case of joint compatible measurements, or to falsify it if the decay rate is incompatible with the observed asymmetries. Having both large values for BR($\mu \to 3e$), and sizeable values for the asymmetries is also a consequence of the new exploration of the parameter space (stepping clearly outside the dipole dominance, and allowing for large values of the Yukawa-like couplings $\mathcal{G}$ and $g_R$). In particular, several operators can be at the origin of significant contributions, leading to the excellent prospects visible in  Fig.~\ref{fig:l3l_Asym_vs_Asym}.

\bigskip
In what concerns the tau 3-body decay asymmetries, and in view of the comparatively poorer prospects to have 
$\tau \to 3\mu$ within experimental reach, the $\mathcal{A}_{T, P, P^\prime}^{\tau \to 3\mu}$ asymmetries can become extraordinarily predictive: phenomenologically allowed points within FCC-ee future reach are characterised by $\mathcal{A}_{T }^{\tau \to 3\mu} \sim 0$, 
$|\mathcal{A}_{P^\prime}^{\tau \to 3\mu}| \leq 0.15$, in association with large $|\mathcal{A}_{P}^{\tau \to 3\mu}|$, up to 90\%.
The asymmetries associated with tau 3-body decays thus emerge as a very powerful probe of this class of models.

Finally, for the final-state mixed flavour cLFV tau decays, one could in principle have extremely large values for all asymmetries; nevertheless, and since the associate decay rates lie outside any future reach, we do not display the results here.

\begin{figure}[h!]
\centering
\includegraphics[width=0.48 \textwidth] {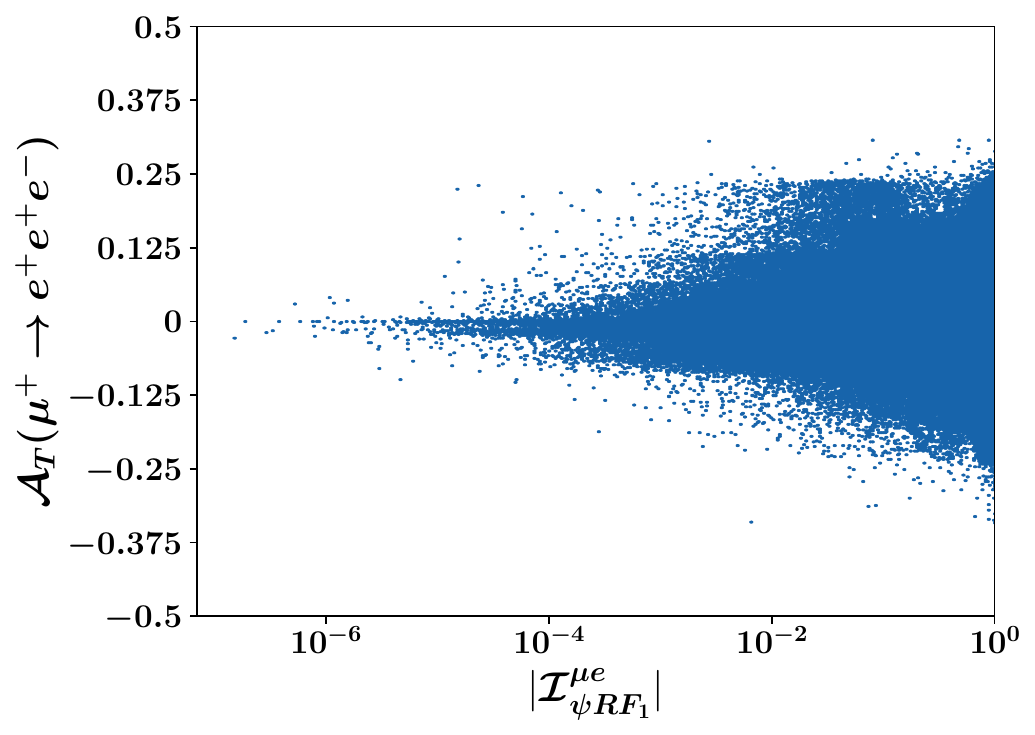}
\caption{Projected values of the $T$ asymmetry for 
$\mu \to 3 e$ decays vs. the invariant $\vert \mathcal{I}_{\psi R F_1}^{\mu e} \vert$ (see Eq.~(\ref{eq:CPV:inv})). 
All displayed points satisfy current experimental constraints.
}
\label{fig:asym_vs_couplings}
\end{figure}

\bigskip
Concerning  $\mathcal{A}_{T }$, which under CPT-invariance probes CP-violation, the complexity of the fermion couplings of the ``T1-2-A'' scotogenic model renders the construction of Jarlskog-like invariants particularly challenging. 
The most promising quantity (amongst those explored) is related to the sextet 
\begin{equation}\label{eq:CPV:inv}
    \mathcal{I}_{I J K}^{\alpha \beta} \equiv \text{Im}\left\{(\hat g_I^{\beta *} \hat g_J^\beta) \,(\hat g_J^{\alpha *} \hat g_K^\alpha)\, (\hat g_K^{\alpha *} \hat g_I^\beta)\right\}\,
\end{equation}
in which $I,J,K \in \{ \psi, F_1, F_2, R\}$, $\alpha, \beta$ are flavour indices and $\hat g \equiv g / \vert g \vert$. Although it is not easy to draw strong implications, we nevertheless point out that, as visible from Fig.~\ref{fig:asym_vs_couplings}, large CP-sensitive $\mathcal{A}_{T }^{\mu \to 3e}$ asymmetries can only occur for sizeable values of $\mathcal{I}_{\psi R F_1}^{\mu e} $ (albeit without any strict dependency nor correlation). Even if it clearly lies outside the scope of the present work, it might be interesting to investigate whether quantities as the above sextet enter the computation of the lepton asymmetries leading to an explanation of the BAU via leptogenesis (see, for instant, the recent works~\cite{Hugle:2018qbw, Alvarez:2023dzz}).

\subsection{Overview}
\begin{figure}[h!]
\centering
\includegraphics[width=0.48 \textwidth] {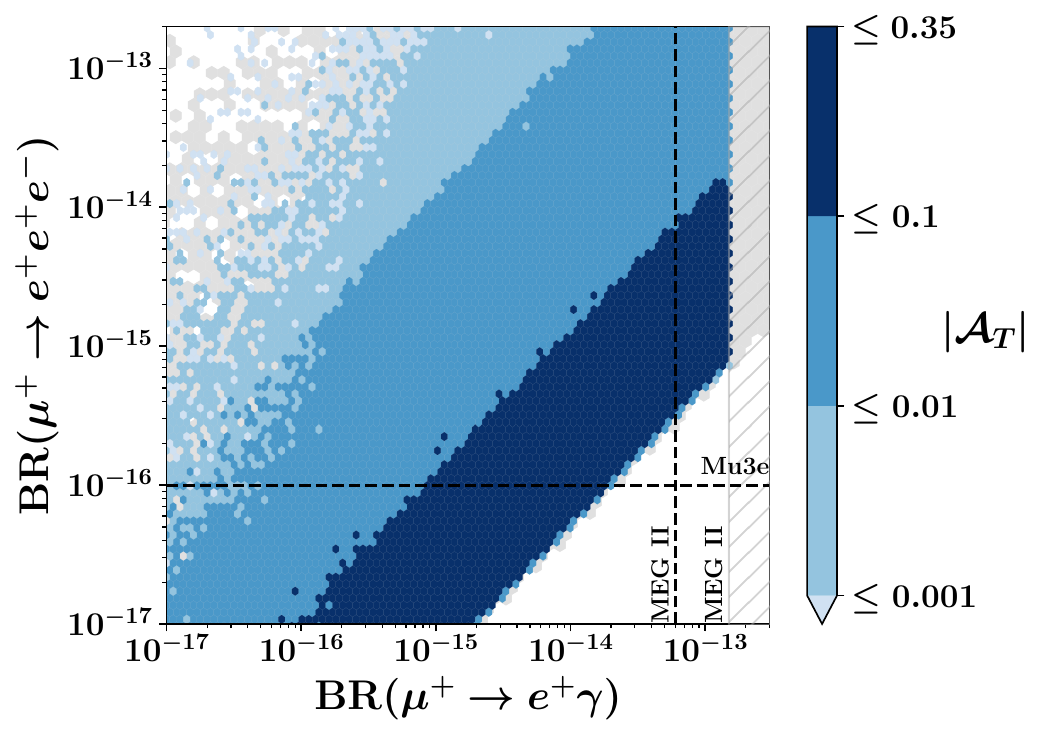}\hspace*{8mm}
\includegraphics[width=0.48 \textwidth] {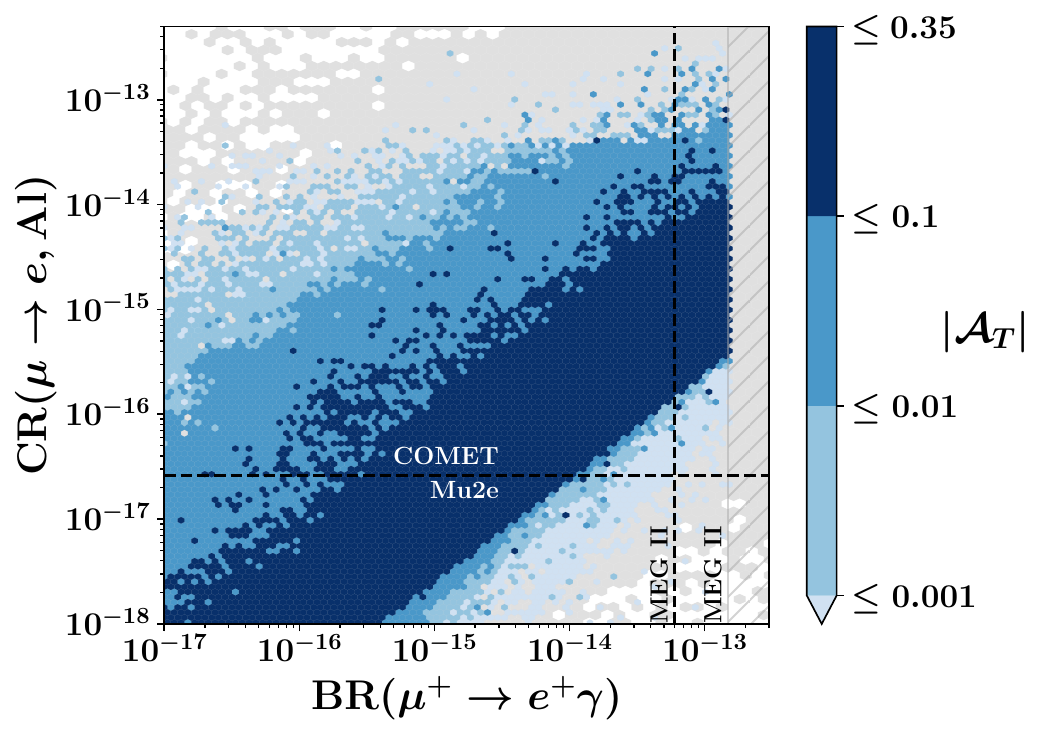}
\caption{Overview of the $T$-asymmetry in association with $\mu \to 3 e$ decays. On the left, projections for $\mathcal{A}_{T }^{\mu \to 3e}$ in the plane spanned by
BR($\mu \to 3e$) and BR($\mu \to e \gamma$); on the right, by CR($\mu-e$, Al) and BR($\mu \to e \gamma$). The blue-coloured bands denote regimes for the maximal possible absolute value of the $T$ asymmetry. Grey and grey-dashed regions correspond to exclusion due to conflict with experimental bounds. Horizontal and vertical lines denote current experimental bounds and future sensitivities.}
\label{fig:Mue_TAsym}
\end{figure}

\begin{figure}[h!]
\centering
\includegraphics[width=0.48 \textwidth] {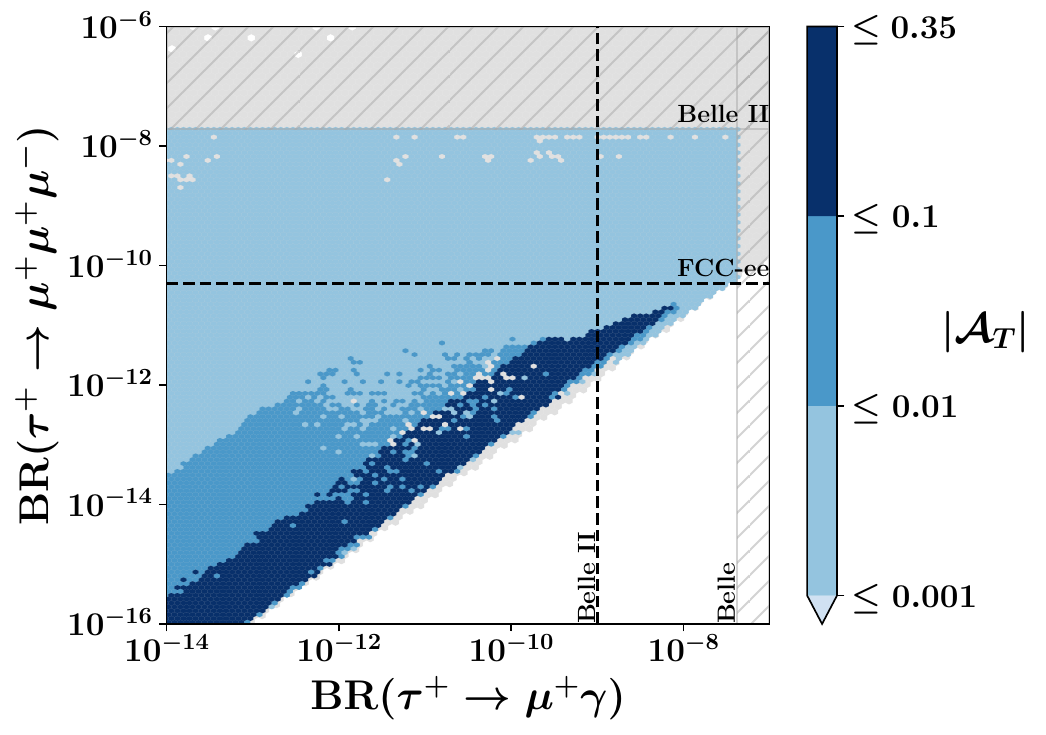}
\caption{Projected values for the $T$-asymmetry 
in association with $\tau \to 3 \mu$ decays, displayed 
in the plane spanned by
BR($\tau \to 3\mu$) and BR($\tau \to \mu \gamma$). Line and colour code as Fig.~\ref{fig:Mue_TAsym}.}
\label{fig:Ta3Mu_TAsym}
\end{figure}

To conclude the study of cLFV observables in the ``T1-2-A'' scotogenic realisation, in what follows we carry out an overview of the joint prospects for the cLFV rates and $T$-asymmetries in 3-body decays. Figures~\ref{fig:Mue_TAsym} and~\ref{fig:Ta3Mu_TAsym} summarise our findings.  

Concerning the $\mu-e$ sector, both panels of Fig.~\ref{fig:Mue_TAsym} offer a comprehensive 
overview of cLFV observables: as can be seen, regimes within MEG II and Mu3e future reach, and with sizeable values for $\mathcal{A}_{T }^{\mu \to 3e}$ (as large as 35\%), are indeed possible in this class of NP models. 
Although the inclusion of the new observables (i.e. the asymmetries) does not translate in new means to readily falsify the model, they nevertheless provide additional important information to characterise the model. Also, a possible sizeable value of $\mathcal{A}_{T }^{\mu \to 3e}$ measured at Mu3e can help to effectively constrain the associated values for neutrinoless conversion in Aluminium (thus expected to lie below $10^{-14}$).

A survey of tau decays into muons is offered in Fig.~\ref{fig:Ta3Mu_TAsym}. We notice that regimes leading to decay rates within FCC-ee future sensitivity are either strongly box dominated or dipole dominated
and feature very small values of $T$-asymmetry (below $1 \%$). Even if 
large asymmetries in this region are in principle possible, they correspond to regions in parameter space already excluded\footnote{We notice that one can indeed find specific cancellations in the Casas-Ibarra parametrisation for the couplings which do lead to regimes within future sensitivities for both the considered cLFV decays (and which would still be compatible with cLFV constraints on muon decays). These are nevertheless statistically disfavoured cases (almost singular) and, in order to pursue such cases, stronger theoretical arguments - as flavour symmetries - would be required.} due to conflict with $\mu$ cLFV decay searches.
Clearly, the present ``T1-2-A'' scotogenic model is not expected to account for any observable asymmetry in 
tau cLFV decays; such an event would thus strongly disfavour this NP scenario.

\section{Outlook}\label{sec:concs}

Scotogenic realisations stand out as a particularly appealing class of SM extensions, in view of their ambitious goal of simultaneously addressing the problem of neutrino mass generation and that of a viable DM candidate. Certain scotogenic models - as the ``T1-2-A'' variant - have become the object of intensive exploration, since they could also explain the baryon asymmetry of the Universe and accommodate tensions between SM prediction and observation, as was the case of the anomalous magnetic moment of the muon.  

Motivated by the proposal to enlarge the set of cLFV observables that might allow probing this class of SM extensions, we have carried out a comprehensive reassessment of the prospects of a ``T1-2-A'' scotogenic variant in what concerns charged lepton flavour violation. In addition to the exploration of a new set of asymmetries associated with cLFV 3-body decays, our study was further fuelled by the impact that the disappearance of the tension in $(g-2)_\mu$ could have in carving out the peculiar cLFV patterns which were intrinsic to this NP model~\cite{Vicente:2014wga, Alvarez:2023dzz, Darricau:2025vcs}.

We have thus performed a sophisticated scan of the model's non-trivial parameter space, relying on a DE-MCMC technique, which allowed bringing to light new features regarding the ``T1-2-A'' scotogenic variant, especially in what concerns cLFV. 
The change in paradigm regarding the anomalous magnetic moment of the muon implies that there is little room left regarding new physics contributions to the flavour conserving dipole operator; this also translates in a departure from the dipole-dominated contributions to cLFV observables which were formerly present. 
We recall that the latter led to significant correlations between radiative decays and 3-body decays (and neutrinoless conversion in nuclei as well), which were proposed as a means to potentially falsify the NP construction in the advent of a cLFV discovery.  

The new comprehensive exploration of the model's parameter space, together with the relaxation of the dipole-dominance, reveals excellent prospects for the observation of an associated cLFV signal. As discussed in the manuscript, and with the exception of  $\tau-e$ modes, 
the contributions of the ``T1-2-A'' scotogenic variant
lead to predictions that are well within reach of the different dedicated facilities, and in many cases already in conflict with current bounds. These regimes are often associated with smaller values of the trilinear coupling $\alpha$ (which were not formerly favoured) and also emerge in association with large values of the Yukawa-like couplings $\mathcal{G}$ and $g_R$. 
For very large values of $\alpha$ - typically above 10~TeV -, one does indeed recover the dipole dominance and the associated correlation between the cLFV observables. 

Enlarging the set of cLFV observables that can be used to test the model thus becomes all the most important. As advocated in previous works~\cite{Goto:2010sn, Bolton:2022lrg, Darricau:2025rmu}, 
several cLFV dedicated experiments offer excellent prospects to measure the angular distributions of 3-body decays, including the $T$, $P$ and $P^\prime$ asymmetries (as well as the forward-backward counterparts). 
Our study suggests that for $\mu \to 3e$ decays one can have large values for the asymmetries: in particular, one can expect up to 25\% for the (CPV-sensitive) 
$\mathcal{A}_{T}$, and larger for $\mathcal{A}_{P,P^\prime}$. 
For the tau sector, in particular in what concerns $\tau \to 3\mu$ decays, points whose cLFV decay rates lie within future experimental reach are in general associated with tiny values of $\mathcal{A}_{T }^{\tau \to 3\mu}$ ($\sim 0$), and large  $\mathcal{A}_{P}^{\tau \to 3\mu}$ (as much as 90\%). This renders  
the tau 3-body decay asymmetries a powerful probe of this class of models, especially in the case of observing $\tau \to 3\mu$ decays. 

While not allowing to compensate for the predictive power of the formerly present correlation of cLFV observables (a consequence of the dominance of dipole contributions), the considered asymmetries do allow to further probe the ``T1-2-A'' scotogenic variant. In the case of $\tau \to 3\mu$ decays, and in view of the experimental prospects, they are  sufficiently predictive  as to potentially falsify the present construction.

Naturally, it would be interesting to explore probes of leptonic CP violation in association with a model which can potentially explain the observed matter-antimatter asymmetry of the Universe via leptogenesis. A next possible step would be to consider if the regimes leading to a viable BAU (see e.g.~\cite{Alvarez:2023dzz}) could also be at the source of sizeable contributions to cLFV observables, in particular the 3-body decays and the associated $T$-asymmetry.

\section*{Acknowledgements}
This project has received support from the IN2P3 (CNRS) Master Project, ``Hunting for Heavy Neutral Leptons'' (12-PH-0100). We are also indebted to E.~Pinsard for critical discussions and advice.

\appendix
\section{Detailed characterisation of the ``T1-2-A'' variant}\label{app:model:description}

Following the addition of new terms to the interaction Lagrangian and scalar potential (as presented in Section~\ref{sec:model}), we review below the enlarged mass spectrum, physical interactions as well as the higher-order contributions to neutrino mass generation. Further details can be found in~\cite{Alvarez:2023dzz, Darricau:2025vcs}.

\subsection{Enlarged spectrum: mass matrices and physical interactions}

\paragraph{Additional fermion states}
While the mass of the charged (Dirac) fermions is given by 
\begin{equation}
m_{\psi^\pm}\, =\, M_{\psi}\, ,
\end{equation}
the physical neutral fermion states are obtained from diagonalising the corresponding mass matrix, given (in the interaction basis) as follows
\begin{equation}\label{eq:chi0:matrix}
M_{\chi^0} \, =\, \begin{pmatrix}
        M_{1} & 0 & \frac{v}{\sqrt{2}}\, y_{11} & \frac{v}{\sqrt{2}} \,y_{21} \\
        0 & M_{2} & \frac{v}{\sqrt{2}}\, y_{12} & \frac{v}{\sqrt{2}} \,y_{22} \\
        \frac{v}{\sqrt{2}} \,y_{11} & \frac{v}{\sqrt{2}} \,y_{12} & 0 & M_{\Psi} \\
        \frac{v}{\sqrt{2}}\, y_{21} & \frac{v}{\sqrt{2}} \,y_{22} & M_{\Psi} & 0 
    \end{pmatrix}\,,
\end{equation}
with the physical and interaction basis related via the unitary mixing matrix $U_{\chi}$:
\begin{equation}\label{eq:chi:M:Uchi}
    \{\chi^0\}^T \, =\, U_\chi^T\, \{ F_1, F_2, \Psi^0_1, (\Psi^0_2)^c \}^T\, 
    \quad \text{in which}\quad 
    M_{\chi^0}\, =\, U_\chi\, M_{\chi^0}^\text{diag}\, U^T_\chi\,.
\end{equation}

\paragraph{Scalar sector}
After EWSB, the scalar sector features the SM-like Higgs (the only neutral scalar developing a vacuum expectation value, $v$), as well as a new doublet and a (real) singlet, \begin{equation}
    H = \begin{pmatrix}
        G^{+} \\
        \frac{1}{\sqrt{2}} \left( v + h^{0} + i G^{0} \right)
    \end{pmatrix}, \quad
    \eta = \begin{pmatrix}
        \eta^{+} \\ \frac{1}{\sqrt{2}} \left( \eta^{0} + i A^{0} \right)
    \end{pmatrix}
    \,,\quad
    S\,.
\end{equation}
Under the assumption of CP-conservation in the scalar sector (which corresponds to choosing $\alpha$ and $\lambda_\eta^{\prime \prime}$ in Eq.~(\ref{eq:Vscalar}) to be real), the scalar sector comprises charged and three neutral states, with no mixing between scalar and pseudoscalar neutral bosons. 
The scalar mass matrix is given by (in the interaction basis)
\begin{equation}\label{MPhi:def}
    M_{\phi}^{2} = \begin{pmatrix}
        M_{S}^{2} + \frac{1}{2} v^{2} \lambda_{S} & v \alpha & 0 \\
        v \alpha & M_{\eta}^{2} + \frac{1}{2} v^{2} \lambda_{L} & 0 \\
        0 & 0 & M_{\eta}^{2} + \frac{1}{2} v^{2} \lambda_{A}
    \end{pmatrix}\,,
\end{equation}
in which $\lambda_{L,A} = \lambda_{\eta} + \lambda_{\eta}^{\prime} \pm \lambda_{\eta}^{\prime\prime}$. The neutral scalar mass matrix can be diagonalised via an unitary mixing matrix $U_{\phi}$ as
\begin{equation}
    \{\phi_1, \phi_2, A_0\}^T \, =\, U_\phi^T\, \{ S_0, \eta_0, A_0 \}^T\, 
    \quad \text{in which}\quad 
    M_{\phi}^2\, =\,U_\phi\, ({M_{\phi}^{\text{diag}}})^2\, U^T_\phi\,,
\end{equation}
with $U_{\phi}$ being parametrised  as 
\begin{equation}\label{eq:UPhi:def}
    U_\phi = 
    \begin{pmatrix}
    \cos{\theta_S} & \sin{\theta_S} & 0\\
    - \sin{\theta_S} & \cos{\theta_S} & 0\\
    0 & 0 & 1\\
    \end{pmatrix}\,,
\end{equation}
and $\theta_S$ given by
\begin{equation}
    \cos{\theta_S} = \frac{\text{sign}(\alpha)}{\sqrt{2}} \sqrt{1 + \frac{ M_\eta^2 - M_S^2 + \frac{1}{2} \left( \lambda_L - \lambda_S \right) v^2 }{\sqrt{4 \alpha^2 v^2 + \left( M_\eta^2 - M_S^2 + \frac{1}{2} \left( \lambda_L - \lambda_S\right) v^2\right)^2}}}\,.
\end{equation}

\subsection{Neutrino mass generation}
Following the approach of previous studies (cf.~\cite{Alvarez:2023dzz, Darricau:2025vcs}), we summarise below the most relevant points leading to the computation of the neutrino masses. 
For convenience, we recall that after EW symmetry breaking, the contributions to the neutrino mass matrix arising from the diagrams of Fig.~\ref{fig:NMassInt} can be written as
\begin{equation}
    \mathcal{M}_{\nu} = \mathcal{G}^T \mathcal{M}_{L} \mathcal{G}, \quad \text{where} \quad \mathcal{G} = \begin{pmatrix}
        g_{\psi}^{e} & g_{\psi}^{\mu} & g_{\psi}^{\tau} \\[5pt]
        g_{F_{1}}^{e} & g_{F_{1}}^{\mu} & g_{F_{1}}^{\tau} \\[5pt]
        g_{F_{2}}^{e} & g_{F_{2}}^{\mu} & g_{F_{2}}^{\tau} 
    \end{pmatrix}\,, \nonumber
\end{equation}
in which $\mathcal{G}$ is a ``coupling'' matrix,
with the $3 \times 3$ symmetric matrix $\mathcal{M}_{L}$ 
encoding the relevant information regarding the new massive fields propagating in the loop~\cite{Alvarez:2023dzz}. The latter can be written in terms of the rotation matrices $U_{\chi}$ and $U_{\phi}$ (see Eqs.~(\ref{eq:chi:M:Uchi}, \ref{eq:UPhi:def})),
\begin{align}
    \left( \mathcal{M}_{L} \right)_{11} &= \sum_{ik} b_{ik} \,( U_{\chi}^* )_{4i}^{2} \,( U_{\phi} )_{1k}^{2}\,, \nonumber\\
    \left( \mathcal{M}_{L} \right)_{22} &= \frac{1}{2} \sum_{ik} b_{ik} \,( U_{\chi}^* )_{1i}^{2} \left[ ( U_{\phi} )_{2k}^{2} - (U_{\phi} )_{3k}^{2} \right] \,,\nonumber\\
    \left( \mathcal{M}_{L} \right)_{33} &= \frac{1}{2} \sum_{ik} b_{ik} \,( U_{\chi}^* )_{2i}^{2} \left[ ( U_{\phi} )_{2k}^{2} - (U_{\phi} )_{3k}^{2} \right] \,,\nonumber\\
    \left( \mathcal{M}_{L} \right)_{12} = \left( \mathcal{M}_{L} \right)_{21} &= \frac{1}{\sqrt{2}} \sum_{ik} b_{ik} \,( U_{\chi}^* )_{1i} \,( U_{\chi}^* )_{4i} \,( U_{\phi} )_{1k} \,( U_{\phi} )_{2k} \,,\nonumber\\
    \left( \mathcal{M}_{L} \right)_{13} = \left( \mathcal{M}_{L} \right)_{31} &= \frac{1}{\sqrt{2}} \sum_{ik} b_{ik}\, ( U_{\chi}^* )_{2i} \,( U_{\chi}^* )_{4i} \,( U_{\phi} )_{1k}\, ( U_{\phi} )_{2k} \,,\nonumber\\
    \left( \mathcal{M}_{L} \right)_{23} = \left( \mathcal{M}_{L} \right)_{32} &= \frac{1}{2} \sum_{ik} b_{ik} \,( U_{\chi}^* )_{2i} \,( U_{\chi}^* )_{1i} \left[ ( U_{\phi} )_{2k}^{2} - ( U_{\phi} )_{3k}^{2} \right]\,.
\end{align}
In the above, $i = 1, 2, 3, 4$ and $k = 1, 2, 3$, and the loop function $b_{ik}$ is given by~\cite{Alvarez:2023dzz}:
\begin{equation}
    b_{ik} \,= \,b(M_{\chi_{_i}},M_{\phi_{_k}}) \,= \,
    \frac{1}{16\pi^2} \,\frac{M_{\chi_{_i}}}{M_{\phi_{_k}}^{2}-M_{\chi_{_i}}^{2}} \left[ M_{\chi_{_i}}^{2} \ln M_{\chi_{_i}}^{2} - M_{\phi_{_k}}^{2} \ln M_{\phi_{_k}}^{2} \right]\,.
\end{equation}
As mentioned in Section~\ref{sec:pheno_impact}, a modified Casas-Ibarra parametrisation~\cite{Casas:2001sr,Basso:2012voo} allows ensuring compatibility with neutrino oscillation data. 
As already presented in Eq.~(\ref{eq:CI-parametrisation}), one has the following parametrisation of $\mathcal{G}$ 
\begin{equation}
    \mathcal{G} \,= U_{L} \,D_{L}^{-1/2}\, R\, D_{\nu}^{1/2} \,U_{\text{PMNS}}^{*}\,,\nonumber
    \label{eqn:casas_ibarra}
\end{equation}
with
\begin{equation}
    D_{L} \,= \,U_{L}^{T} \,\mathcal{M}_{L} \,U_{L}\,.\nonumber
\end{equation}
We recall that in the above $D_{\nu}$ is the diagonal matrix of ligth neutrino masses, and the unitary $3 \times 3$ $U_{\text{PMNS}}$ matrix encodes the leptonic mixing data; the $R$ mixing matrix can be parametrised as
\begin{equation}
    R = \begin{pmatrix}
        c_3 & -s_3 & 0\\
        s_3 & c_3 & 0\\
        0 & 0 & 1
    \end{pmatrix}\begin{pmatrix}
        c_1 & 0 & -s_1\\
        0 & 1 & 0\\
        s_1 & 0 & c_1
    \end{pmatrix}\begin{pmatrix}
        1 & 0 & 0\\
        0 & c_2 & -s_2\\
        0 & s_2 & c_2
    \end{pmatrix}\,,
 \label{eqn:R_matrix}
\end{equation}
where $s_{i} = \sin\theta_{i}$ and $c_{i} = \cos\theta_{i}$, with complex $\theta_{1,2,3}$.

\section{Leptonic cLFV observables}\label{app:cLFV-details}
In this appendix we provide the relevant elements to the computation of the leptonic cLFV observables which play a prominent role in our study - radiative and three-body decays, and $\mu-e$ conversion in nuclei. Other cLFV decays which only play a constraining role in assessing the viability of the parameter space (as is the case of the $Z$- and Higgs-boson decays)
and further observables including EW precision tests, have been addressed in~\cite{Darricau:2025vcs}, in which the details for the computation can be found.

\subsection{Form factors}
\paragraph{Radiative decays}
Following~\cite{Darricau:2025vcs}, the rates of the cLFV radiative decays can be expressed as
\begin{equation}
        \text{BR}( \ell_{\alpha} \rightarrow \ell_{\beta} \gamma ) \,= \,\frac{m_{\ell_{\alpha}}^{3}}{4\pi\Gamma_{\ell_{\alpha}}} \left(|c_{R}^{\alpha\beta} |^{2} + | c_{R}^{\beta\alpha} |^{2} \right)\,,
    \label{eqn:cLFV:radiative}
\end{equation}
in which $\Gamma_{\ell_{\alpha}}$ is the charged lepton decay width. In the above, 
the coefficient of the dipole contribution can be given by~\cite{Alvarez:2023dzz}
\begin{eqnarray}\label{eqn:wilson_cRij}
    c_{R}^{\alpha \beta} & = & \sum_i \frac{e}{64 \pi^2 M_{\chi_{_i}}^2} \left[ (\Gamma^{\beta i}_L )^*\, \Gamma^{\alpha i}_R\, M_{\chi_{_i}}\, f_{\chi,2}^\gamma(x_i) + \left( m_{\ell_\beta} ( \Gamma^{\beta i}_L )^*\, \Gamma^{\alpha i}_L + m_{\ell_\alpha} ( \Gamma^{\beta i}_R )^* \,\Gamma^{\alpha i}_R \right) g_2^\gamma(x_i) \right] \nonumber \\
    & - & \sum_k \frac{e}{64 \pi^2 M_{\phi_{_k}}^2} \left[ ( \Gamma^{\beta k}_L )^*\, \Gamma^{\alpha k}_R \,M_{\phi_{_k}} \,f_{\phi,2}^\gamma(x_k) + \left( m_{\ell_\beta} ( \Gamma^{\beta k}_L )^* \,\Gamma^{\alpha k}_L + m_{\ell_\alpha}( \Gamma^{\beta k}_R )^* \, \Gamma^{\alpha k}_R \right) g_2^\gamma(x_k) \right],
\end{eqnarray}
in which $f_2^\gamma$ and $g_2^\gamma$ are the loop functions given in Appendix~\ref{app:loop_func}, $x_i = M_{\eta^\pm}^2/M_{\chi_{_i}}^2$, $x_k = M_\psi^2/M_{\phi_k}^2$ denote the mass ratios of the new particles in the loop, $e$ is the electric charge, $m_{\ell_\alpha}$ denotes the charged lepton mass, and $\Gamma_{L,R}$ the left- and right-handed interactions between physical states; for the present NP construction, the latter are respectively given by
\begin{align}\label{eq:GammaLR}
    \Gamma_L^{\alpha k}  &=  - g_\psi^\alpha\, U_\phi^{1 k}\,, \nonumber \\
    \Gamma_R^{\alpha k} & = \frac{\left(g_R^\alpha \right)^*}{\sqrt{2}} 
 \left( U_\phi^{2 k} + i U_\phi^{3 k} \right)\,, \nonumber \\ 
 \Gamma_L^{\alpha i} & =  g_{F_1}^\alpha \left(U_\chi^{1 i} \right)^* + g_{F_2}^\alpha \left(U_\chi^{2 i} \right)^* \,, \nonumber \\
    \Gamma_R^{\alpha i} & = \left(g_R^\alpha \right)^* 
 U_\chi^{3 i}\,.
\end{align}

\paragraph{Three-body decays}
The cLFV 3-body decays receive contributions from several distinct operators, including photon-, $Z$- and Higgs-penguins, as well as box diagrams. In the following, we adopt the notation and the effective operator basis of~\cite{Abada:2014kba}.
We recall that the dipole form factors (i.e. $K_{2,L}^{\alpha \beta} = 2 c_R^{\alpha \beta}/(e \, m_{\ell_{_\alpha}})$) were already given in Eq.~(\ref{eqn:wilson_cRij}). The anapole form factors can be written as:
\begin{align}\label{eq:anapoleFF}
    K_{1,L}^{\alpha \beta} &= \frac{1}{576 \pi^2} \left[ \sum_i \Gamma^{\alpha i}_L \left(\Gamma^{\beta i}_L\right)^* \frac{f_{\chi, 1}^\gamma (x_i)}{M_{\chi_{_i}}^2} + \sum_k \Gamma^{\alpha k}_L \left(\Gamma^{\beta k}_L\right)^* \frac{f_{\phi, 1}^\gamma (x_k)}{M_{\phi_{_k}}^2} \right]\, , \nonumber 
    \\
    K_{1,R}^{\alpha \beta} &= K_{1,L}^{\alpha \beta} \quad (L \leftrightarrow R)\, ,
\end{align}
in which $f_1^\gamma$ are the loop functions (see Appendix~\ref{app:loop_func}). The $Z$-penguin form factors are:
\begin{align}
    F_{Z,L}^{\alpha \beta} &= - \frac{e}{128 \pi^2 c_w s_w} \left[\sum_{i j} \Gamma^{\alpha i}_L \left(\Gamma^{\beta j}_L\right)^* \left( \Gamma_Z^{i j} \, g_2^Z(x_i, x_j) + \left(\Gamma_Z^{i j}\right)^* f_{2 \chi}^Z(x_i, x_j)\right) \right]\, , \nonumber 
    \\
    F_{Z,R}^{\alpha \beta} &= \frac{e}{128 \pi^2 c_w s_w} \left[\sum_i \Gamma^{\alpha i}_R \left( \left(\Gamma^{\beta i}_R\right)^* f_{1 \chi}^Z (x_i) + \sum_j \left(\Gamma^{\beta j}_R\right)^* \left( \Gamma_Z^{i j} \, f_{2 \chi}^Z(x_i, x_j) + \left(\Gamma_Z^{i j}\right)^* \left(g_2^Z(x_i, x_j) - 1 \right)\right) \right) \right.\, \nonumber\\
    & + \left. \sum_k \left(\Gamma^{\alpha k}_R \left(\Gamma^{\beta k}_R\right)^* f_{1 \phi}^Z (x_k) + i U_\phi^{2 k} \left( \Gamma^{\alpha k}_R \left(\Gamma^{\beta 3}_R\right)^* - \Gamma^{\alpha 3}_R \left(\Gamma^{\beta k}_R\right)^*\right) \left(g_2^Z(x_3, x_k) - 3 \right) \right)\right] \,,
\end{align}
where $f^Z$ and $g^Z$ are loop functions given in Appendix~\ref{app:loop_func}, and $\Gamma_Z^{i j} = U_\chi^{3 i} ( U_\chi^{3 j})^* - U_\chi^{4 i} ( U_\chi^{4 j})^*$. The Higgs-penguin form factors are given by:
\begin{align}
    F_{H,L}^{\alpha \beta} &= \frac{1}{16 \pi^2} \left[ \sum_i \Gamma^{\alpha i}_L \left( \left(\Gamma^{\beta i}_R\right)^* M_{\chi_{_i}} \left( \frac{\lambda_\eta v}{M_{\eta^\pm}^2}f_{1 \chi}^H (x_i) + \frac{1}{v} g_1^H(x_i) \right) \right. \right. \nonumber \\
    & + \left. \left. \sum_j \left(\Gamma^{\beta j}_R\right)^* \left( \Gamma_{H,L}^{i j} \, \left(g_2^Z(x_i, x_j) + 1 \right) + \left(\Gamma_{H,L}^{i j}\right)^* f_{2 \chi}^H(x_i, x_j)\right) \right)\right.\, \nonumber\\
    & + \left. \sum_k \left(\Gamma^{\beta k}_R\right)^* \left(\Gamma^{\alpha k}_L g_1^H (x_k) + \sum_l \Gamma^{\alpha l}_L \Gamma_H^{k l} f_{2\phi}^H(x_k,x_l) \right) \right] \,, \nonumber\\
    F_{H,R}^{\alpha \beta} &= F_{H,L}^{\alpha \beta} (L \leftrightarrow R) \, .
\end{align}
Again the loop functions ($f^H$ and $g^H$) can be found in Appendix~\ref{app:loop_func}, and
\begin{align}\label{eq:HVert}
    \Gamma_{H,L}^{i j} & = \sum_{a b} \frac{y_{a b}}{\sqrt{2}} \,\left( U_\chi^{a + 2, i}\, U_\chi^{b, j} + U_\chi^{a + 2, i}\, U_\chi^{b, j} \right)\,, \\
    \Gamma_{H,R}^{i j} & = ( \Gamma_{H,L}^{i j})^*\,,\\
    \Gamma_H^{k l} & = - \left( \left[ v \,\lambda_\eta^{\prime \prime} \,
    \left( U_\phi^{2 k} + i U_\phi^{3 k}\right) \left( U_\phi^{2 l} + i U_\phi^{3 l}\right) + 2\, \alpha \,U_\phi^{1 k} U_\phi^{2 l} + \text{c.c.}(k \leftrightarrow l)\right] \right. \nonumber \\
    & + \left. v 
    \left[ \lambda_S \,U_\phi^{1 k} \,U_\phi^{1 l} + \left( \lambda_\eta + \lambda_\eta^\prime \right) \left( U_\phi^{2 k} + i U_\phi^{3 k}\right) \left( U_\phi^{2 l} - i U_\phi^{3 l}\right) + (k \leftrightarrow l)\right]\right)\,.
\end{align}
The box form factors can be written as:
\begin{align}
    B_{S,LL}^{\alpha \beta \gamma \delta} &= \frac{1}{128 \pi^2} \left[ \frac{1}{M_{\eta^\pm}^2}\sum_{i j} \Gamma_L^{\alpha i} \left( \Gamma_L^{\delta j} \left( \Gamma_R^{\beta j} \Gamma_R^{\gamma i} - 2 \Gamma_R^{\beta i} \Gamma_R^{\gamma j} \right)^* + 2 \Gamma_L^{\delta i} \left( \Gamma_R^{\beta j} \Gamma_R^{\gamma j} \right)^* \right) f^{\text{Box}}_{2 \chi, SXX} (x_i, x_j) \right.\nonumber\\
    &+ \left. \frac{1}{M_\psi^2}\sum_{k l} \Gamma_L^{\alpha k} \left( \Gamma_L^{\delta l} \left( \Gamma_R^{\beta k} \Gamma_R^{\gamma l} - 2 \Gamma_R^{\beta l} \Gamma_R^{\gamma k} \right)^* + 2 \Gamma_L^{\delta k} \left( \Gamma_R^{\beta l} \Gamma_R^{\gamma l} \right)^* \right) f^{\text{Box}}_{2 \phi, SXX} (x_k, x_l) \right]\, , \nonumber
    \\
    B_{S,RR}^{\alpha \beta \gamma \delta} &= B_{S,LL}^{\alpha \beta \gamma \delta} (L \leftrightarrow R) \, , \nonumber 
    \\
    B_{S,LR}^{\alpha \beta \gamma \delta} &= \frac{1}{128 \pi^2} \left[\frac{1}{M_{\eta^\pm}^2}\sum_{i j} \Gamma_L^{\alpha i} \Gamma_R^{\delta j} \left( \left( \Gamma_R^{\beta j} \Gamma_L^{\gamma i} \right)^* f^{\text{Box}}_{2 \chi, SXY} (x_i, x_j) + \left( \Gamma_R^{\beta i} \Gamma_L^{\gamma j} \right)^* g^{\text{Box}}_{2 \chi, SXY} (x_i, x_j) \right) \right.\nonumber\\
    &+ \left. \frac{1}{M_{\psi}^2} \sum_{k l} \Gamma_L^{\alpha k} \left( \Gamma_R^{\delta l} \left( \Gamma_R^{\beta k} \Gamma_L^{\gamma l} - 2 \Gamma_R^{\beta l} \Gamma_L^{\gamma k} \right)^* - 3 \Gamma_L^{\delta k} \left( \Gamma_R^{\beta l} \Gamma_L^{\gamma l} \right)^* f^{\text{Box}}_{2 \phi, SXY} (x_k, x_l) \right. \right. \nonumber \\
    &+ \left. \left. \Gamma_R^{\delta l} \left( \Gamma_R^{\beta k} \Gamma_L^{\gamma l} + \Gamma_R^{\beta l} \Gamma_L^{\gamma k} \right)^* g^{\text{Box}}_{2 \phi, SXY} (x_k, x_l)\right) \right] \, , \nonumber
    \\
    B_{S,RL}^{\alpha \beta \gamma \delta} &= B_{S,LR}^{\alpha \beta \gamma \delta} (L \leftrightarrow R) \, , \nonumber 
    \\
    B_{V,LL}^{\alpha \beta \gamma \delta} &= \frac{1}{128 \pi^2} \left[ \frac{1}{M_{\eta^\pm}^2}\sum_{i j} \Gamma_L^{\alpha i} \left( \Gamma_L^{\delta j} \left(  \Gamma_L^{\beta j} \Gamma_L^{\gamma i} + \Gamma_L^{\beta i} \Gamma_L^{\gamma j}  \right)^* f^{\text{Box}}_{2 \chi, VXX} (x_i, x_j) + \Gamma_L^{\delta i} \left( \Gamma_R^{\beta j} \Gamma_L^{\gamma j} \right)^* g^{\text{Box}}_{2 \chi, VXX} (x_i, x_j) \right) \right. \nonumber\\
    &+ \left. \frac{1}{M_{\psi}^2}\sum_{k l} \Gamma_L^{\alpha k} \left( \Gamma_L^{\delta l} \left( \Gamma_L^{\beta k} \Gamma_L^{\gamma l} + \Gamma_L^{\beta l} \Gamma_L^{\gamma k} \right)^* - 2 \Gamma_L^{\delta k} \left( \Gamma_L^{\beta l} \Gamma_L^{\gamma l} \right)^* \right) f^{\text{Box}}_{2 \phi, VXX} (x_k, x_l) \right] \, , \nonumber
    \\
    B_{V,RR}^{\alpha \beta \gamma \delta} &= B_{V,LL}^{\alpha \beta \gamma \delta} (L \leftrightarrow R) \, , \nonumber 
    \\
    B_{V,LR}^{\alpha \beta \gamma \delta} &= \frac{1}{128 \pi^2} \left[ \frac{1}{M_{\eta^\pm}^2}\sum_{i j} \Gamma_L^{\alpha i} \Gamma_R^{\delta j} \left( \left( \Gamma_L^{\beta i} \Gamma_R^{\gamma j} \right)^* f^{\text{Box}}_{2 \chi, VXY} (x_i, x_j) + \left( \Gamma_L^{\beta j} \Gamma_R^{\gamma i} \right)^* g^{\text{Box}}_{2 \chi, VXY} (x_i, x_j) \right) \right. \nonumber\\
    &+ \left. \frac{1}{M_{\psi}^2}\sum_{k l} \Gamma_L^{\alpha k} \left( \Gamma_R^{\delta l} \left( 2 \Gamma_L^{\beta k} \Gamma_R^{\gamma l} - \Gamma_L^{\beta l} \Gamma_R^{\gamma k} \right)^* + 3 \Gamma_R^{\delta k} \left( \Gamma_L^{\beta l} \Gamma_R^{\gamma l} \right)^* f^{\text{Box}}_{2 \phi, VXY} (x_k, x_l) \right. \right. \nonumber \\
    &+ \left. \left. \Gamma_R^{\delta l} \left( \Gamma_L^{\beta k} \Gamma_R^{\gamma l} + \Gamma_L^{\beta l} \Gamma_R^{\gamma k} \right)^* g^{\text{Box}}_{2 \phi, VXY} (x_k, x_l)\right) \right]\, , \nonumber
    \\
    B_{V,RL}^{\alpha \beta \gamma \delta} &= B_{V,LR}^{\alpha \beta \gamma \delta} (L \leftrightarrow R) \, , \nonumber 
    \\
    B_{T,LL}^{\alpha \beta \gamma \delta} &= \frac{1}{128 \pi^2} \left[ \frac{1}{M_{\eta^\pm}^2}\sum_{i j} \Gamma_L^{\alpha i} \left( \Gamma_L^{\delta j} \left( \Gamma_R^{\beta j} \Gamma_R^{\gamma i}\right)^* - 2 \Gamma_L^{\delta i} \left( \Gamma_R^{\beta j} \Gamma_R^{\gamma j} \right)^* \right) f^{\text{Box}}_{2 \chi, TXX} (x_i, x_j) \right. \nonumber\\
    &+ \left. \frac{1}{M_{\psi}^2}\sum_{k l} \Gamma_L^{\alpha k} \left( \Gamma_L^{\delta l} \left( \Gamma_R^{\beta k} \Gamma_R^{\gamma l}\right)^* + \Gamma_L^{\delta k} \left( \Gamma_R^{\beta l} \Gamma_R^{\gamma l} \right)^* \right) f^{\text{Box}}_{2 \phi, TXX} (x_k, x_l) \right]\, , \nonumber
    \\
    B_{T,RR}^{\alpha \beta \gamma \delta} &= B_{T,LL}^{\alpha \beta \gamma \delta} (L \leftrightarrow R) \, ,
\end{align}
in which  $f^\text{Box}$ and $g^\text{Box}$ are loop functions (see Appendix~\ref{app:loop_func}). Finally, by matching the box and penguin contributions to the $4 \ell$ operators we obtain the following Wilson coefficients:
\begin{align}
    A_{S, XY}^{\alpha \beta \beta \beta} &= \frac{1}{2} B_{S, XY}^{\alpha \beta \beta \beta} + \frac{F_{H,Y}^{\alpha \beta} R_{H,X}^{\beta \beta}}{M_H^2} \, , \quad A_{S, XY}^{\alpha \beta \gamma \gamma} = B_{S, XY}^{\alpha \beta \gamma \gamma} + \frac{F_{H,Y}^{\alpha \beta} R_{H,X}^{\gamma \gamma}}{M_H^2} \, , \quad A_{S, XY}^{\alpha \beta \beta \gamma} = \frac{1}{2} B_{S, XY}^{\alpha \beta \beta \gamma} \, , \nonumber 
    \\
    A_{V, XY}^{\alpha \beta \beta \beta} &= \frac{1}{2} B_{V, XY}^{\alpha \beta \beta \beta} + \frac{F_{Z,Y}^{\alpha \beta} R_{Z,X}^{\beta \beta}}{M_Z^2} + e^2 K_{1,Y}^{\alpha \beta} \, , \quad A_{V, XY}^{\alpha \beta \gamma \gamma} = B_{V, XY}^{\alpha \beta \gamma \gamma} + \frac{F_{Z,Y}^{\alpha \beta} R_{Z,X}^{\gamma \gamma}}{M_Z^2} + e^2 K_{1,Y}^{\alpha \beta} \, , \quad A_{V, XY}^{\alpha \beta \beta \gamma} = \frac{1}{2} B_{V, XY}^{\alpha \beta \beta \gamma}, \nonumber 
    \\
    A_{T, XX}^{\alpha \beta \beta \beta} &= \frac{1}{2} B_{T, XX}^{\alpha \beta \beta \beta} \, , \quad A_{T, XX}^{\alpha \beta \gamma \gamma} = B_{T, XX}^{\alpha \beta \gamma \gamma} \, , \quad A_{T, XX}^{\alpha \beta \beta \gamma} = \frac{1}{2} B_{T, XX}^{\alpha \beta \beta \gamma} \, ,
\end{align}
where $\alpha \neq \beta \neq \gamma$, $X,Y \in \{L,R\}$, $e$ is the electric charge and $R_{Z,L}^{\ell \ell} = - e (c_w^2 - s_w^2)/(2 c_w s_w) \,, R_{Z,R}^{\ell \ell} = e s_w/c_w \, , R_{H,L}^{\alpha \alpha}=R_{H,R}^{\alpha \alpha}= - m_{\ell_{_\alpha}}/v$ are the tree level lepton couplings to the $Z$ and Higgs respectively.

\paragraph{Muon-electron conversion in nuclei} We use the expressions of~\cite{Kitano:2002mt} to compute the neutrinoless $\mu-e$ conversion in nuclei. The conversion rate can be cast as:
\begin{align}
    \text{CR}\left( \mu \rightarrow e, \text{Nucleus}\left(D,S^{(p)},V^{(p)},S^{(n)},V^{(n)}\right) \right) &= \frac{32 G_F^2 m_\mu^5}{\Gamma_\text{capt}} \left( \vert \mathcal{A}_L \vert^2 + \vert \mathcal{A}_R \vert^2 \right) \, ,
\end{align}
where $G_F = e^2/(4 \sqrt{2} s_w^2 M_W^2) $ is the Fermi constant, $\Gamma_\text{capt}$ is the muon capture rate, $D,S^{(p)},V^{(p)},S^{(n)},V^{(n)}$ are the dipole, proton and neutron scalar/vector overlap integrals respectively, whose values for the different atoms are given in~\cite{Kitano:2002mt}. The amplitudes $\mathcal{A}_{L/R}$ can be cast as:
\begin{align}
    \mathcal{A}_L &= A_L \frac{D}{4} + g_{L,S}^{(p)} S^{(p)} + g_{L,V}^{(p)} V^{(p)} + g_{L,S}^{(n)} S^{(n)} + g_{L,V}^{(n)} V^{(n)} \, , \nonumber \\
    \mathcal{A}_R &= \mathcal{A}_L (L \leftrightarrow R) \, ,
\end{align}
in which
\begin{align}
    g_{X,I}^{(p)} &= \sum_q g_{X,I}^{(q)} G_I^{(q,p)} \, , \nonumber
    \\
    g_{X,I}^{(n)} &= \sum_q g_{X,I}^{(q)} G_I^{(q,n)} \, , \nonumber
\end{align}
with $I \in \{ S, V\}$, $X \in \{ L , R \}$. The nucleon form factors $G_K$~\cite{Kosmas:2001mv} are
\begin{eqnarray}
    G_V^{(u,p)} = G_V^{(d,n)} = 2\,, \quad
   & G_V^{(d,p)} = G_V^{(u,n)} = 1\,, \quad
   & G_V^{(s,p)} = G_V^{(s,n)} = 0\,,
    \nonumber\\
   G_S^{(u,p)} = G_S^{(d,n)} = 5.1\,, \quad
 &   G_S^{(d,p)} = G_S^{(u,n)} = 4.3\,, \quad
  &  G_S^{(s,p)} = G_S^{(s,n)} = 2.5\,.
\end{eqnarray}
Finally, the effective couplings can be expressed in terms of the underlying form factors as:
\begin{align}
    A_L &= \frac{e}{4 \sqrt{2} G_F} K_{2,L}^{\mu e} \, , \nonumber
    \\
    A_R &= A_L (L \leftrightarrow R) \, , \nonumber
    \\
    g_{L,S}^{(q)} &= -\frac{1}{\sqrt{2} G_F} \frac{R_{H}^{qq} F_{H,L}^{\mu e}}{M_H^2} \, , \nonumber 
    \\
    g_{R,S}^{(q)} &= g_{L,S}^{(q)} (L \leftrightarrow R) \, , \nonumber
    \\
    g_{L,V}^{(q)} &= -\frac{1}{2 \sqrt{2} G_F} \left( Q_q e K_{1,L}^{\mu e} + \frac{R_{Z}^{qq} F_{Z,L}^{\mu e}}{M_Z^2} \right) \, , \nonumber 
    \\
    g_{R,V}^{(q)} &= g_{L,V}^{(q)} (L \leftrightarrow R) \, ,
\end{align}
in which $R_{Z}^{u u} = - e (2 s_w/(3 c_w) + (3 c_w^2 - s_w^2)/(6 c_w s_w)) \,, R_{Z}^{d d} = - e (s_w/(3 c_w) + (3 c_w^2 + s_w^2)/(6 c_w s_w)) \, , R_{H}^{q q}= - m_q/v$ are the vector-like tree level quark couplings to the $Z$ and Higgs respectively, and $Q_q, m_q$ are the electric charge and mass of the quarks in the nuclei respectively.

\subsection{Loop functions}\label{app:loop_func}
The loop functions associated to the dipole form-factors are:
\begin{align}
       f_{\chi,2}^\gamma(x) &= \frac{x^{2} - 1 - 2x\ln (x)}{
       (x-1)^{3}}\, , \quad \forall x \neq 1\, ;
       \quad 
        f_{\chi,2}^\gamma(1) = \frac{1}{3} \,,\nonumber 
        \\
       f_{\phi,2}^\gamma(x) &= \frac{x^{2} - 4x + 3 + 2\ln (x)}{
       (x-1)^{3}}\, , \quad 
       \forall x \neq 1\, ;
       \quad 
        f_{\phi,2}^\gamma(1) = \frac{2}{3} \,,\nonumber 
        \\
        g_2^\gamma(x) &= \frac{x^{3} - 6x^{2} + 3x + 2 + 6x\ln (x)}{(x-1)^{4}}
        \, , \quad 
        \forall x \neq 1\, ;
       \quad 
        g_2^\gamma(1) = \frac{1}{12} \,. 
\end{align}
The loop functions associated to the anapole form factors are given by
\begin{align}
    f_{\chi,1}^\gamma(x) &= \frac{2 x^2 - 7x + 11}{(x-1)^3} - \frac{6 \ln (x)}{(x-1)^4}\, , \quad \forall x \neq 1\, ; \quad f_{\chi,1}^\gamma(1) = \frac{3}{2}\,,\nonumber 
    \\
    f_{\phi,1}^\gamma(x) &= \frac{7 x^2 - 29 x + 16}{(x-1)^3} + \frac{6 (3 x - 2) \ln (x)}{(x-1)^4}\, , \quad \forall x \neq 1\, ; \quad f_{\phi,1}^\gamma(1) = \frac{9}{2}\,.
\end{align}
Concerning the $Z$-penguin contribution, the associated loop functions can be cast as
\begin{align}
    f_{1\phi}^Z(x) &= \frac{3 x - 1}{x-1} - \frac{2 (2 x - 1) \ln (x)}{(x-1)^2}\,,\nonumber 
    \\
    f_{1\chi}^Z(x) &= \frac{x - 3}{x-1} + \frac{2 \ln (x)}{(x-1)^2}\,,\nonumber 
    \\
    f_{2\chi}^Z(x,y) &= 4 \frac{\sqrt{x y}}{x - y} \left(\frac{\ln (y)}{y-1} - \frac{\ln(x)}{x-1}\right)\,, \nonumber 
    \\
    g_2^Z(x,y) &= \frac{2}{x - y} \left(\frac{x \ln (y)}{y-1} - \frac{y \ln(x)}{x-1}\right)\,.
\end{align}
Likewise, the loop functions associated to the Higgs-penguin contributions are:
\begin{align}
    f_{1\chi}^H(x) &= - \frac{x}{x-1} + \frac{x \ln (x)}{(x-1)^2}\,,\nonumber 
    \\
    g_1^H(x) &= \frac{\ln (x)}{x-1},\nonumber 
    \\
    f_{2\chi}^H(x,y) &= - \frac{1}{4} f_{2\chi}^Z(x,y)\,, \nonumber 
    \\
    f_{2\phi}^H(x,y) &= \frac{\sqrt{x y}}{4} f_{2\chi}^Z(x,y)\,, \nonumber 
    \\
    g_2^H(x,y) &= -\frac{1}{2} g_2^Z(x,y)\,.
\end{align}
Finally, the loop functions associated to box diagram contributions are given by
\begin{align}
    f_{2\phi,SXX}^{\text{Box}}(x,y) &= 4 x y \left( \frac{1}{(x-1)(y-1)} + \frac{1}{x-y} \left( \frac{x \ln(x)}{(x-1)^2} - \frac{y \ln(y)}{(y-1)^2}\right) \right)\,,\nonumber 
    \\
    f_{2\phi,SXY}^{\text{Box}}(x,y) &= 4 x y \left( \frac{1}{(x-1)(y-1)} + \frac{1}{3(x-y)} \left( \frac{(2 x + 1) \ln(x)}{(x-1)^2} - \frac{(2 y + 1) \ln(y)}{(y-1)^2}\right) \right)\,,\nonumber 
    \\
    g_{2\phi,SXY}^{\text{Box}}(x,y) &= \frac{2 \sqrt{x y}}{3} f_{2\chi}^Z(x,y)\,,\nonumber 
    \\
    f_{2\phi,VXX}^{\text{Box}}(x,y) &= -2 x y \left( \frac{1}{(x-1)(y-1)} + \frac{1}{x-y} \left( \frac{\ln(x)}{(x-1)^2} - \frac{\ln(y)}{(y-1)^2}\right) \right)\,,\nonumber 
    \\
    f_{2\phi,VXY}^{\text{Box}}(x,y) &= \frac{1}{2} f_{2\phi,SXY}^{\text{Box}}(x,y)\,,\nonumber 
    \\
    g_{2\phi,VXY}^{\text{Box}}(x,y) &= - \frac{1}{2} g_{2\phi,SXY}^{\text{Box}}(x,y)\,,\nonumber 
    \\
    f_{2\phi,TXX}^{\text{Box}}(x,y) &= \frac{1}{4} f_{2\phi,SXX}^{\text{Box}}(x,y)\,,\nonumber 
    \\
    f_{2\chi,SXX}^{\text{Box}}(x,y) &= \frac{1}{\sqrt{x y}} f_{2\phi,SXX}^{\text{Box}}(x,y)\,,\nonumber 
    \\
    f_{2\chi,SXY}^{\text{Box}}(x,y) &= -2 f_{2\phi,VXX}^{\text{Box}}(x,y)\,,\nonumber 
    \\
    g_{2\chi,SXY}^{\text{Box}}(x,y) &= -\frac{2}{\sqrt{x y}} f_{2\chi,SXX}^{\text{Box}}(x,y)\,,\nonumber 
    \\
    f_{2\chi,VXX}^{\text{Box}}(x,y) &= f_{2\phi,VXX}^{\text{Box}}(x,y)\,,\nonumber 
    \\
    g_{2\chi,VXX}^{\text{Box}}(x,y) &= -\frac{2}{\sqrt{x y}} f_{2\chi,SXX}^{\text{Box}}(x,y)\,,\nonumber 
    \\
    f_{2\chi,VXY}^{\text{Box}}(x,y) &= f_{2\phi,VXX}^{\text{Box}}(x,y)\,,\nonumber 
    \\
    g_{2\chi,VXY}^{\text{Box}}(x,y) &= \frac{1}{\sqrt{x y}} f_{2\chi,SXX}^{\text{Box}}(x,y)\,,\nonumber 
    \\
    f_{2\chi,TXX}^{\text{Box}}(x,y) &= \frac{1}{4 \sqrt{x y}} f_{2\chi,SXX}^{\text{Box}}(x,y)\,.
\end{align}

{
\small
\bibliographystyle{JHEP}
\bibliography{bibliography}
}

\end{document}